\documentclass[11pt,american,twoside]{article}
\usepackage[T1]{fontenc}
\usepackage[utf8]{inputenc}
\usepackage{latexsym}
\usepackage{amsmath}
\usepackage{bbm}
\usepackage{amssymb}
\usepackage{babel}
\usepackage{amsfonts}
\usepackage{fourier}
\usepackage{ntheorem}
\usepackage{epsfig}
\usepackage{enumerate}
\usepackage{color,xcolor}
\usepackage{xspace,needspace,xparse,easyReview}
\usepackage[active]{srcltx}
\usepackage[colorlinks=true]{hyperref}
\usepackage{fancyhdr}
\usepackage{xargs}
\usepackage[toc]{appendix}
\usepackage[normalem]{ulem}
\colorlet{darkblue}{blue!50!black}
\hypersetup{
	colorlinks,%
	citecolor=darkblue,%
	filecolor=red,%
	linkcolor=darkblue,%
	urlcolor=darkblue,%
	pdfnewwindow=true,%
	pdfstartview={FitH}
}

\numberwithin{equation}{section}

\newcounter{smallarabics}

\newcounter{smallroman}

\newcommand{\ben}{\begin{enumerate}[{\rm (1)}]}
\newcommand{\een}{\end{enumerate}}

\newtheorem{theorem}{Theorem}[section]
\newtheorem{proposition}[theorem]{Proposition}
\newtheorem{lemma}[theorem]{Lemma}
\newtheorem{corollary}[theorem]{Corollary}
\theorembodyfont{\upshape}
\newtheorem{definition}[theorem]{Definition}
\newtheorem{remark}[theorem]{Remark}

\AtBeginEnvironment{theorem}{\Needspace{5\baselineskip}}
\AtBeginEnvironment{proposition}{\Needspace{5\baselineskip}}
\AtBeginEnvironment{definition}{\Needspace{5\baselineskip}}
\AtBeginEnvironment{corollary}{\Needspace{5\baselineskip}}
\AtBeginEnvironment{lemma}{\Needspace{5\baselineskip}}
\AtBeginEnvironment{quote}{\Needspace{5\baselineskip}}
\newcommand{\beq}{\begin{equation}}
\newcommand{\eeq}{\end{equation}}
\newcommand{\bear}[1]{\begin{array}{#1}}
\newcommand{\ear}{\end{array}}
\newcommand{\bep}{\begin{proposition}}
\newcommand{\eep}{\end{proposition}}
\newcommand{\bet}{\begin{theoreme}}
\newcommand{\eet}{\end{theoreme}}
\newcommand{\bel}{\begin{lemma}}
\newcommand{\eel}{\end{lemma}}
\newcommand{\bed}{\begin{definition}}
\newcommand{\eed}{\end{definition}}
\newcommand{\ber}{\begin{remark}}
\newcommand{\eer}{\end{remark}}
\newcommand{\proof}{\noindent{\bf Proof.}~}
\newcommand{\qed}{\hfill$\Box$}

\NewDocumentCommand{\ASS}{mm}{\expandafter\newcommand\csname #1\endcsname{{\hyperref[#1]{\bf (#2)}}}}
\NewDocumentCommand{\preASS}{mm}{\expandafter\newcommand\csname pre#1\endcsname{{\hyperref[#1]{\bf (#2)}}}}
\newcommand{\assuming}[3]{
\begin{quote}\label{#2}{\bf(#1) }%
#3%
\end{quote}%
\ASS{#2}{#1}}

\newcommand{\ds}{\displaystyle}
\newcommand{\twol}[2]{\genfrac{}{}{0pt}{1}{#1}{#2}}

\def\cA{\mathcal{A}} \def\cB{\mathcal{B}} \def\cC{\mathcal{C}}
 \def\cE{\mathcal{E}} \def\cF{\mathcal{F}}
 \def\cH{\mathcal{H}} 
 \def\cK{\mathcal{K}} \def\cL{\mathcal{L}}
 \def\cN{\mathcal{N}} \def\cO{\mathcal{O}}
\def\cP{\mathcal{P}} \def\cQ{\mathcal{Q}} 
\def\cS{\mathcal{S}}

\def\NN{\mathbb{N}}

\def\CC{\mathbb{C}}

\def\II{\mathbb{I}}

\def\RR{\mathbb{R}}
\def\one{\mathbb{1}}

\def\fh{\mathfrak{h}}
\def\fH{\mathfrak{H}}
\def\fI{\mathfrak{I}}
\def\fM{\mathfrak{M}}
\def\fF{\mathfrak{F}}
\def\fG{\mathfrak{G}}

\def\bF{\boldsymbol{F}}

\def\sS{\mathsf{S}}
\def\sR{\mathsf{R}}

\newcommand{\slim}{\mathop{\mathrm{s-lim}}\limits}

\newcommand{\tr}{\mathop{\mathrm{tr}}\nolimits}

\newcommand{\ran}{\mathop{\mathrm{Ran}}}
\renewcommand{\ker}{\mathop{\mathrm{Ker}}}
\newcommand{\Dom}{\mathop{\mathrm{Dom}}}
\renewcommand{\sp}{\mathop{\mathrm{sp}}}
\renewcommand{\Re}{\mathop{\mathrm{Re}}}
\renewcommand{\Im}{\mathop{\mathrm{Im}}}
\newcommand{\dist}{\mathop{\mathrm{dist}}}

\newcommand{\e}{\mathrm{e}}
\renewcommand{\d}{\mathrm{d}}
\renewcommand{\i}{\mathrm{i}}

\newcommand{\BAR}{\overline}

\renewcommand{\dim}{\mathop{\mathrm{dim}}}

\newcommand{\CAR}{\mathop{\mathrm{CAR}}}

\newcommand{\tphi}{\widetilde{\varphi}}
\newcommand{\ttm}{\mathrm{2tm}}
\newcommand{\anc}{\mathrm{ancilla}}
\newcommand{\qpsc}{\mathrm{qpsc}}
\newcommand{\fr}{\mathrm{fr}}
\newcommand{\Ent}{\mathop{\mathrm{Ent}}}

\newcommand{\AnV}{{\bf (AnV$\boldsymbol{(\vartheta)}$)}}

\begin{document}
\def\today{}
\title{Entropic Fluctuation Theorems for the Spin--Fermion Model}
\author{T. Benoist$^{1}$, L. Bruneau$^{2}$, V. Jak\v{s}i\'c$^{3}$, A. Panati$^4$, C.-A. Pillet$^4$
\\ \\
$^1$ Institut de Mathématiques de Toulouse, UMR5219,\\
 Université de Toulouse, CNRS, UPS IMT, F-31062 Toulouse Cedex 9, France\\ \\
$^2$ D\'epartement de Math\'ematiques, CNRS UMR 8088\\
Cergy Paris Université, 95000 Cergy-Pontoise, France
\\ \\
$^3$Dipartimento di Matematica\\
Politecnico di Milano\\
piazza Leonardo da Vinci, 32 \\
20133 Milano,  Italy
\\ \\
$^4$Universit\'e de Toulon, CNRS, CPT, UMR 7332, 83957 La Garde, France\\
Aix-Marseille Universit\'e, CNRS, CPT, UMR 7332, Case 907, 13288 Marseille, France
}
\maketitle
\thispagestyle{empty}

\noindent{\small{\bf Abstract.}
We study entropic fluctuations in the Spin--Fermion model describing an $N$-level quantum system coupled to several independent thermal free Fermi gas reservoirs.
We establish the quantum Evans--Searles and Gallavotti--Cohen fluctuation theorems and identify their link with entropic ancilla state tomography and quantum phase space contraction of non-equilibrium steady state.
The method of proof involves the spectral resonance theory of quantum transfer operators developed by the authors in previous works.
}
\tableofcontents

\section{Introduction}
\label{sec:introduction}

This is the fourth and final paper in a
series~\cite{Benoist2023a,Benoist2024b,Benoist2024} dealing with entropic
fluctuations in quantum statistical mechanics, and in particular with the
quantum Evans--Searles and Gallavotti--Cohen fluctuation theorems.  Its goal is
to illustrate, on the example of the open spin--fermion model, the general
theory developed in~\cite{Benoist2023a,Benoist2024}. The
work~\cite{Benoist2024b} was devoted to the justification of the key formulas
of~\cite{Benoist2023a} by thermodynamic limit arguments.

We assume the reader to be familiar with the framework and results
of~\cite{Benoist2023a,Benoist2024b,Benoist2024}. In particular, we will use the
notation and conventions regarding open quantum systems and modular theory
introduced in these works.

In the context of open quantum systems, the spin--fermion model goes back to the
works~\cite{Davies1974,Spohn1978b}, and with time has become one of the
paradigmatic models of quantum statistical mechanics. The closely related
spin--boson model, in which each thermal reservoir is a free Bose gas, has a
much longer history in the physics literature due to its connection with
non-relativistic QED; see e.g.~\cite[Section~1.6]{Derezinski2001}. The
description and analysis of the spin--boson model is technically more involved,
and the model does not fit directly in the $C^\ast$-algebraic formalism
of~\cite{Benoist2023a,Benoist2024}.

The revival of interest in the spin--fermion/boson model started
with~\cite{Jaksic1996b,Jaksic1996a,Bach1998} that have generated a large body of
literature; an incomplete list of references
is~\cite{Huebner1995,Derezinski1999,Bach2000,Derezinski2001,Merkli2001,
Froehlich2002,Jaksic2002a,Derezinski2003,Froehlich2004,Jaksic2006a,
Merkli2007a,Salem2007,Derezinski2008a,DeRoeck2009,DeRoeck2013,Moller2014,
Hasler2021}. We will comment on some of these works as we proceed.
The techniques we will use draw on~\cite{Jaksic1996b,Jaksic1996a,
Jaksic2002a,Jaksic2010b}.

The paper is organized as follows. In Section~\ref{sec:sfmodel}, we introduce the
spin--fermion model, briefly recall the main objects of study and state our
results. In Section~\ref{sec:qto}, for the convenience of the reader, we recall
the modular structure of the model, as well as the $\alpha$-Liouvilleans
introduced in~\cite{Jaksic2010b,Benoist2024} and their connection with the
various entropic functionals. Section~\ref{sec:deformedliouvillean} is devoted
to the study of these $\alpha$-Liouvilleans and closely follows the analysis
in~\cite{Jaksic1996b, Jaksic1996a, Jaksic2002a}. The proof of the main theorem
is given in Section~\ref{sec:proof-mainthm}.

\paragraph*{Acknowledgments} The work of CAP and VJ was partly funded by the
CY Initiative grant {\sl Investissements d'Avenir}, grant number ANR-16-IDEX-0008.
The work of TB was funded by the ANR project {\sl ESQuisses}, grant number
ANR-20-CE47-0014-01, and by the ANR project {\sl Quantum Trajectories}, grant
number ANR-20-CE40-0024-01. VJ acknowledges the support of NSERC  and the support of the MUR
grant "Dipartimento di Eccellenza 2023-2027" of Dipartimento di Matematica, Politecnico di Milano. A part of this
work was done during long term visits of LB and AP to McGill and CRM--CNRS
International Research Laboratory IRL 3457 at University of Montreal. The LB
visit was funded by the CNRS and AP visits by the CRM Simons and
FRQNT--CRM--CNRS programs. We also acknowledge the support of
the ANR project {\sl DYNACQUS}, grant number ANR-24-CE40-5714.

\section{The Spin--Fermion Model}
\label{sec:sfmodel}

\subsection{Description of the model}
\label{ssec:model}

The spin--fermion model is a concrete example of open quantum system with the
structure described in~\cite[Section~1.1]{Benoist2023a}, where several
independent reservoirs are coupled through a small system $\sS$. The model has a
non-trivial small system part, described by a finite dimensional Hilbert space
$\cK_\sS$ and Hamiltonian $H_\sS$. Its $C^\ast$-algebra of observables is
$\cO_\sS=\cB(\cK_\sS)$, where $\cB(\cH)$ denotes the $C^\ast$-algebra of all
bounded operators on a Hilbert space $\cH$. Its dynamics
$\tau_\sS^t=\e^{t\delta_\sS}$ is generated by $\delta_\sS=\i[H_\sS,\,\cdot\;]$
and its reference state is $\omega_\sS(A)=\tr(A)/\dim\,\cK_\sS$\footnote{This
choice is made for convenience. None of our results depend on the choice of
$\omega_\sS$ as long as $\omega_\sS>0$.}. Each reservoir subsystem $\sR_j$,
$1\leq j\leq M$, is a free Fermi gas with single particle Hilbert space $\fh_j$
and single particle Hamiltonian $h_j$. The algebra of observables of $\sR_j$ is
the CAR-algebra $\cO_j=\CAR(\fh_j)$, the $C^\ast$-algebra generated by
creation/annihilation operators $a_j^\ast(f)$/$a_j(f)$, $f\in\fh_j$, satisfying
the canonical anti-commutation relations
$$
\{a_j(f),a_j^\ast(g)\}=\langle f,g\rangle\one,\qquad
\{a_j(f),a_j(g)\}=0.
$$
The Heisenberg dynamics on $\cO_j$ is the group of Bogoliubov
$\ast$-auto\-morphisms associated to $h_j$, {\sl i.e.,} the $C^\ast$-dynamics
defined by $\tau_j^t(a_j(f))=a_j\left(\e^{\i th_j}f\right)$. We denote by
$\delta_j$ its generator, $\tau_j^t=\e^{t\delta_j}$. The reference state
$\omega_j$ on $\cO_j$ is the gauge-invariant quasi-free state generated by the
Fermi-Dirac density operator
\[
T_j=\left(\one+\e^{\beta_jh_j}\right)^{-1},
\]
where $\beta_j>0$ is the inverse temperature. $\omega_j$ is the unique
$(\tau_j,\beta_j)$-KMS state on $\cO_j$. The  full reservoir system
$\sR=\sR_1+\cdots+\sR_M$ is described by  the quantum dynamical system
$(\cO_\sR,\tau_\sR,\omega_\sR)$ where
\[
\begin{split}
\cO_\sR&=\cO_1\otimes\cdots\otimes \cO_M,\\[1mm]
\tau_\sR&=\tau_1\otimes\cdots\otimes\tau_M,\\[1mm]
\omega_\sR&=\omega_1\otimes \cdots \otimes \omega_M.
\end{split}
\]
The $C^\ast$-algebra and reference state of the joint system $\sS+\sR$ are
$\cO=\cO_\sS\otimes\cO_\sR$ and $\omega=\omega_\sS\otimes\omega_\sR$. In the
absence of interaction between $\sS$ and $\sR$, its dynamics is
$\tau_\fr=\tau_\sS\otimes\tau_\sR$. This free dynamics is generated by
$\delta_\fr=\delta_\sS+\delta_1+\cdots+\delta_M$.\footnote{Whenever the meaning
is clear within the context,  we write $A$ for $A\otimes \one$ and $\one \otimes
A$, $\delta_j$ for $\delta_j\otimes {\rm Id}$, ${\rm Id}\otimes \delta_j$, etc.}

For each $j$, the interaction of $\sS$ with $\sR_j$ is described by
\beq\label{sf-form}
V_j=\sum_{k=1}^{m_j}Q_{j,k}\otimes R_{j,k}\in\cO_\sS\otimes\cO_j,
\eeq
where
$Q_{j,k}\in\cO_\sS$ is self-adjoint and
\beq\label{sf-form-1}
R_{j,k}=\i^{n_{j,k}(n_{j,k }-1)/2}\varphi_j(f_{j,k,1})\cdots\varphi_j(f_{j,k,n_{j,k}}),
\eeq
with form factors $f_{j,k,m}\in\fh_j$, and where
$\varphi_j(f)=\frac1{\sqrt2}(a_j(f)+a_j^\ast(f))$ are the Segal field operators.
Following~\cite{Davies1974}, we assume that:

\assuming{SFM0}{SFMzero}{For all $t\in\RR$, $j\in\{1,\ldots,M\}$ and
$(k,m)\not=(k^\prime,m^\prime)$,
\[
\langle f_{j,k,m},\e^{\i th_j}f_{j,k^\prime,m^\prime}\rangle=0.
\]
}
In particular, taking $t=0$ in~\SFMzero{} ensures that the $R_{j,k}$ are
self-adjoint elements of $\cO_j$.

Without further mentioning we will assume~\SFMzero{} throughout the paper.
The complete interaction is $V=\sum_j V_j$, and the interacting dynamics
$\tau_\lambda$ is generated by
\[
\delta=\delta_\fr+\lambda\i[V,\,\cdot\;],
\]
where $\lambda\in\RR$ is a coupling constant. The coupled system $\sS+\sR$
is described by the $C^\ast$-quantum dynamical system $(\cO,\tau_\lambda,\omega)$.
We denote by $\omega_t=\omega\circ\tau_\lambda^t$ the evolution of
the state $\omega$ at time $t$.

\ber In the simplest and most studied example of spin--fermion model one has
$\cK_\sS=\CC^2$, $H_\sS=\sigma_z$ and for each reservoir $\sR_j$ the interaction
is of the form $V_j=\sigma_x\otimes \varphi(f_j)$, where $\sigma_z$ and
$\sigma_x$ denote the usual Pauli matrices, see the example at the end of
Section~\ref{ssec:main-results}.
\eer

\medskip
As already mentioned, we are interested in the quantum versions of both
Evans--Searles and Galla\-votti--Cohen fluctuation theorems, a convenient
reference in the spirit of the present  work is the review~\cite{Jaksic2011}.
The former refers to entropic fluctuations with respect to the initial
(reference) state $\omega$ of the system while the latter refers to these
fluctuations with respect to the Non-Equilibrium Steady State (NESS)
$\omega_+$ of the system.  The next assumption postulates the existence of
such an NESS of $(\cO,\tau_\lambda,\omega)$.

\assuming{SFM1}{SFMone}{For all $A\in\cO$ the limit
\[
\omega_+(A)=\lim_{t\uparrow\infty}\omega_t(A)
\]
exists, and the restriction $\omega_{+\sS}$ of the state $\omega_+$ to
$\cO_\sS$ is faithful, $\omega_{+\sS}>0$.
}

In Section~\ref{ssec:main-results} we will describe sufficient conditions that
guarantee the validity of~\SFMone.

A time reversal of the $C^\ast$-dynamics $\tau_\lambda$ is an anti-linear
involutive $\ast$-automorphism $\Theta$ of $\cO$ such that
$\Theta\circ\tau_\lambda^t=\tau_\lambda^{-t}\circ\Theta$ for all $t\in\RR$. A
state $\nu$ on $(\cO,\tau_\lambda)$ is time-reversal invariant if $\tau_\lambda$
admits a time reversal $\Theta$ such that $\nu\circ\Theta(A)=\nu(A^\ast)$ for
all $A\in\cO$. In this case, we will say that the quantum dynamical system
$(\cO,\tau_\lambda,\nu)$ is time-reversal invariant (TRI).

If $(\cO,\tau_\lambda,\omega)$ is TRI, then~\SFMone{} implies that for all
$A\in\cO$ the limit
\[
\omega_-(A)=\lim_{t\to-\infty}\omega_t(A)
\]
exists, and is given by $\omega_-(A)=\omega_+\circ\Theta(A^\ast)$.

\subsection{Entropy production and entropic functionals}
\label{ssec:entropicfunctionals}

Before introducing the three entropic functionals which are the main objects of
our study, we briefly recall the mathematical framework needed to define these
objects. The purpose here is to fix our notation, and we must refer the reader
to~\cite{Benoist2023a,Benoist2024} for a more detailed introduction and
discussions.

Let $(\cH,\pi,\Omega)$ be the GNS-representation of $\cO$ induced by $\omega$.
The {\sl enveloping algebra\/} $\fM$ is the smallest von~Neumann subalgebra of
$\cB(\cH)$ containing $\pi(\cO)$. A state on $\fM$ is {\sl normal} whenever it
is described by a density matrix on $\cH$. The states on $\cO$ obtained as
restrictions of these normal states are called $\omega$-{\sl normal} and form
the {\sl folium} $\cN$ of $\omega$.

The dynamical system $(\cO,\tau_\lambda,\omega)$ is {\sl modular:} $\omega$ is a
$(\varsigma_\omega,-1)$-KMS state where
$\varsigma_\omega^t=\e^{t\delta_\omega}$, the {\sl modular group} of $\omega$,
is the $C^\ast$-dynamics generated by \[ \delta_\omega=-\sum_{j=1}^M \beta_j
\delta_j. \] Since $\delta_j(\varphi_j(f))=\varphi_j(\i h_jf)$, our next
assumption ensures that $V_j\in\Dom(\delta_j)$, and hence
$V\in\Dom(\delta_\omega)$.

\assuming{SFM2}{SFMtwo}{$f_{j,k,m}\in\Dom(h_j)$ for all $j,k,m$.}

The observable $\Phi_j=-\lambda\delta_j(V_j)$ is then well-defined, and
describes the {\sl energy flux} out of the $j^\text{th}$ reservoir. This brings
us to the notion of {\sl entropy production rate,} given by the
observable~\cite{Jaksic2001a,Ruelle2001}
$$
\sigma=\lambda\delta_\omega(V)=\sum_{j=1}^M \beta_j \Phi_j,
$$
satisfying the {\sl entropy balance relation,} see e.g.~\cite{Jaksic2001a},
\beq
\Ent(\omega_t|\omega)=-\int_0^t\omega_s(\sigma)\d s.
\label{ep-eq}
\eeq
The left-hand side of this relation is Araki's {\sl relative
entropy}~\cite{Araki1975/76,Araki1977}, with the sign and ordering convention
of~\cite{Jaksic2001a}.  Since this quantity is non-positive, one has
$\int_0^t\omega_s(\sigma)\d s\geq0$ for all $t\in\RR$, and hence
$\omega_+(\sigma)\geq0$.

\ber Whenever $\Theta$ is a time reversal for $\tau_\lambda$, irrespective of
the coupling $\lambda\in\RR$, then $\Theta(V)=V$ and $\Theta(\sigma)=-\sigma$,
so that $\omega_-(\sigma)\le0$.  In particular, $\omega_+(\sigma)=0$ if
$\omega_-=\omega_+$.
\eer

For $\alpha\in\i\RR$, the {\sl Connes cocycle} of a pair $(\mu,\nu)$ of
faithful $\omega$-normal states is
\[
[D\mu:D\nu]_{\alpha}=\Delta_{\mu|\nu}^\alpha\Delta_{\nu}^{-\alpha},
\]
where $\Delta_\nu$, the {\sl modular operator} of the state $\nu$, and
$\Delta_{\mu|\nu}$, the {\sl relative modular operator} of the  pair
$(\mu,\nu)$, are both non-negative operators on $\cH$.  Thus,
$([D\mu:D\nu]_\alpha)_{\alpha\in\i\RR}$ is a family of unitaries on $\cH$ which,
in fact, belong to $\fM$, see e.g.~\cite[Appendix B]{Araki1982}. We further have

\bep\label{prop:sfm2-reg2}
Suppose \SFMtwo{} holds. Then, for all $t\in\RR$,
$([D\omega_t:D\omega]_{\alpha})_{\alpha\in\i\RR}\subset\pi(\cO)$.
\eep

In what follows we write $[D\omega_t:D\omega]_\alpha$ for
$\pi^{-1}([D\omega_t:D\omega]_{\alpha})$. Similarly, whenever the meaning is
clear within the context, we write $A$ for $\pi(A)$. The proof of the above
proposition relies on the identity
\beq
\log\Delta_{\omega_t|\omega}=\log\Delta_\omega+Q_t,\qquad
Q_t=\int_0^t\tau_\lambda^{-s}(\sigma)\d s,
\label{eq:sunnybutcold}
\eeq
and the subsequent norm convergent expansion
\beq
[D\omega_t:D\omega]_\alpha=\one+\sum_{n\geq1}\alpha^n \int\limits_{0\leq\theta_1\leq\cdots\leq\theta_n\leq1} \varsigma_{\omega}^{-\i\theta_1\alpha}(Q_t)\cdots \varsigma_\omega^{-\i\theta_n\alpha}(Q_t)\d\theta_1\cdots\d\theta_n.
\label{eq:cocycle-dyson}
\eeq
For more details about
Relations~\eqref{eq:sunnybutcold}--\eqref{eq:cocycle-dyson}, we refer the reader
to~\cite[Section~2]{Benoist2024b}, and in particular Lemma~2.4 and
Equation~(2.13) in this reference.

We can now introduce the three entropic functionals considered
in~\cite{Benoist2024}. We only briefly recall their definition and refer the
reader to~\cite{Benoist2023a,Benoist2024} for an in depth discussion.

\subsubsection*{Two-time measurement entropy production (2TMEP)}

The following result was established in~\cite[Theorem~1.3]{Benoist2023a}.%
\footnote{Assumption~({\bf Reg2}) of~\cite{Benoist2023a} is guaranteed
by Proposition~\ref{prop:sfm2-reg2}}
For any $\nu\in\cN$, $t\in\RR$, and $\alpha\in\i\RR$, the limit
$$
\fF_{\nu,t}^\ttm(\alpha)=\lim_{R\to\infty}\frac1R\int_0^R
\nu\circ\varsigma_\omega^\theta\left([D\omega_{-t}:D\omega]_{\alpha}\right)\d\theta
$$
exists, and there is a unique Borel probability measure $Q_{\nu,t}^\ttm$
on $\RR$ such that
\beq
\fF_{\nu,t}^\ttm(\alpha)=\int_\RR\e^{-\alpha s}\d Q_{\nu,t}^\ttm(s).
\label{def:ttm-measure}
\eeq
Moreover, one also has that
$$
\fF_{\nu,t}^\ttm(\alpha)=\lim_{R\to\infty}\frac1R\int_0^R\nu\circ\varsigma_\omega^\theta
\left([D\omega_{-t}:D\omega]_{\frac{\bar \alpha}{2}}^\ast
[D\omega_{-t}:D\omega]_{\frac{\alpha}{2}}\right)\d\theta.
$$
As discussed in~\cite{Benoist2023a,Benoist2024}, the family
$\left(Q_{\nu,t}^\ttm\right)_{t\in\RR}$ describes the statistics of a two-time
measurement of the entropy produced during a time period of length $t$ in the
system $(\cO,\tau_\lambda,\omega)$, when the latter is in the state $\nu$ at the
time of the first measurement. In~\cite{Benoist2023a} it was also shown that, if
each reservoir system $(\cO_j,\tau_j,\omega_j)$ is ergodic,\footnote{This holds
if the one-particle Hamiltonian $h_j$ has purely absolutely continuous spectrum,
see~\cite{Pillet2006} for a pedagogical discussion of this topic.} then the map
\[
\cN\ni\nu\mapsto Q_{\nu,t}\in\cP(\RR),
\]
where $\cP(\RR)$ denotes the set of all Borel probability measures on $\RR$
equipped with the weak topology, extends by continuity to the set $\cS_\cO$ of
all states on $\cO$, equipped with the weak$^\ast$ topology. This defines
$Q_{\nu,t}$, hence $\fF_{\nu,t}^\ttm$ by~\eqref{def:ttm-measure}, for all
$\nu\in\cS_\cO$. We will be particularly interested in the case $\nu=\omega_+$.

\subsubsection*{Entropic ancilla state tomography (EAST)}

For $\nu\in\cS_\cO$, $t\in\RR$ and $\alpha\in\i\RR$, we set
$$
\fF_{\nu,t}^\anc(\alpha)=\nu\left([D\omega_{-t}:D\omega]^\ast_{\frac{\bar\alpha}{2}}
[D\omega_{-t}:D\omega]_{\frac{\alpha}{2}}\right).
$$
{\bf EAST} is described by the family of functionals
$\left(\fF_{\nu,t}^\anc\right)_{t\in\RR}$. When $\nu=\omega$, and up to an
irrelevant prefactor, $\fF_{\omega,t}^\anc(\alpha)$ provides an experimental
implementation of  $\fF_{\omega,t}^\ttm(\alpha)$ through coupling and specific
indirect projective measurements on an ancilla, a procedure called ancilla state
tomography, see~\cite[Section~2.4]{Benoist2024} and references therein.

\subsubsection*{Quantum phase space contraction (QPSC)}

For $\nu\in\cS_\cO$, $t\in\RR$ and $\alpha\in\i\RR$, we set
$$
\fF_{ \nu,t}^\qpsc(\alpha)=\nu\left([D\omega_{-t}:D\omega]_{\alpha}\right).
$$
{\bf QPSC} is described by the family of functionals
$\left(\fF_{\nu,t}^\qpsc\right)_{t\in\RR}$ and provides another
natural route to the quantization of the classical entropic
functionals~\cite[Section 2.7]{Benoist2024}.

\medskip
Note that when $\nu=\omega$ the three families of functionals coincide,
\beq
\fF_{\omega,t}^\ttm=\fF_{\omega,t}^\anc=\fF_{\omega,t}^\qpsc,
\label{eq:functionals-equality}
\eeq
and that
\beq
\partial_\alpha\fF_{\omega,t}^\ttm(\alpha)\big|_{\alpha=0}
=-\int_\RR s\,\d Q_{\omega,t}^\ttm(s)=\Ent(\omega_t|\omega).
\label{eq:functional-derivative}
\eeq
The equalities~\eqref{eq:functionals-equality} are however broken
if $\omega$ is replaced by some other state $\nu\in\cS_\cO$.

\subsection{Fluctuation theorems and the principle of regular entropic fluctuations}
\label{ssec:fluctuation-thm}

We first strengthen \SFMtwo{} to

\assuming{SFM3}{SFMthree}{
$f_{j,k,m}\in\Dom\left(\e^{a|h_j|}\right)$ for all $a>0$ and all $j,k,m$.
}

Since $\varsigma_\omega^\theta(\varphi_j(f))=\varphi\left(\e^{-\i\theta\beta_jh_j}f\right)$,
Assumption~\SFMthree{} guarantees that $V$ is an entire element for the modular group
$\varsigma_\omega$, so that the regularity assumption~\AnV{} of~\cite{Benoist2024}
is satisfied for any $\vartheta>0$.

Our next assumption ensures that all the reservoir subsystems
$(\cO_j,\tau_j,\omega_j)$ are ergodic. As a consequence, the probability
distribution $Q_{\omega_+,t}^\ttm$ and entropic functional
$\fF^\ttm_{\omega_+,t}$ are well-defined.

\assuming{SFM4}{SFMfour}{
$h_j$ has purely absolutely continuous spectrum for all $j$.
}

The last two assumptions have the following consequence
\begin{theorem}
Suppose that \SFMone, \SFMthree{} and \SFMfour{} hold.  Then, for all $t\in\RR$:
\ben
\item The map
\[
\i\RR\ni\alpha\mapsto[D\omega_t:D\omega]_\alpha\in\cO
\]
extends to an entire analytic function.
\item The maps
\[
\begin{split}
\alpha &\mapsto\fF_{\omega,t}^\ttm(\alpha),\\[1mm]
\alpha &\mapsto\fF_{\omega_+,t}^\ttm(\alpha),\\[1mm]
\alpha &\mapsto\fF_{\omega_+,t}^\anc(\alpha),\\[1mm]
\alpha &\mapsto\fF_{\omega_+,t}^\qpsc(\alpha),
\end{split}
\]
defined for $\alpha\in\i\RR$, extend to entire analytic functions. Moreover, for all $\alpha\in\CC$,
\[
\fF_{\omega,t}^\ttm(\alpha)=\int_\RR\e^{-\alpha s}\d Q_{\omega,t}^\ttm(s),\qquad
\fF_{\omega_+,t}^\ttm(\alpha)=\int_\RR\e^{-\alpha s}\d Q_{\omega_+,t}^\ttm(s).
\]
\item The measures $Q_{\omega,t}^\ttm$ and $Q_{\omega_+,t}^\ttm$ are equivalent,
\footnote{They have the same sets of measure zero.}
and for some constants $k,K>0$ and all $t\in\RR$,
\[
k\leq\frac{\d Q_{\omega_+,t}^\ttm}{\d Q_{\omega,t}^\ttm}\leq K.
\]
\een
\label{prelim-thm}
\end{theorem}

\proof Part~(1) follows from~\SFMthree{} and~\cite[Proposition~2.11]{Benoist2024} while
Part~(2) is a consequence of~\SFMone+\SFMthree, \cite[Proposition~3.2]{Benoist2024}
and a well known property of Laplace transforms. Finally, Part~(3) follows
from~\SFMfour{} and~\cite[Theorem~1.6]{Benoist2023a}.
\hfill\qed

\medskip
Assuming that \SFMone--\SFMfour{} hold, we are now ready to introduce the
principle of regular entropic fluctuations (abbreviated the PREF)
of~\cite{Benoist2024}. There, the PREF was introduced in several forms: {\sl
weak, strong,} and {\sl strong + qpsc.} Here we will deal only with the latest
(and strongest) form, and therefore we drop its qualification.

\begin{definition}\label{def-PREF}
Let $I=\,]\vartheta_-,\vartheta_+[$ be an open interval containing $0$.
We say that $(\cO,\tau_\lambda,\omega)$ satisfies the PREF on the interval
$I$ if, for all $\alpha\in I$, the limits
\beq
\begin{split}
\bF^\ttm_{\omega}(\alpha)&=\lim_{t\to\infty}\frac1t\log\fF_{\omega,t}^\ttm(\alpha),\\[1mm]
\bF^\ttm_{\omega_+}(\alpha)&=\lim_{t\to\infty}\frac1t\log\fF_{\omega_+,t}^\ttm(\alpha),\\[1mm]
\bF^\anc_{\omega_+}(\alpha)&=\lim_{t\to\infty}\frac1t\log\fF^\anc_{\omega_+,t}(\alpha),\\[1mm]
\bF^\qpsc_{\omega_+}(\alpha)&=\lim_{t\to\infty}\frac1t\log\fF^\qpsc_{\omega_+,t}(\alpha)
\end{split}
\label{def:functionals-limit}
\eeq
exist, and define differentiable functions on $I$, satisfying
\beq
\bF^\ttm_{\omega}=\bF^\ttm_{\omega_+}=\bF^{\rm ancilla}_{\omega_+}=\bF^\qpsc_{\omega_+}.
\label{def:pref}
\eeq
We denote by $\bF$ the common function in~\eqref{def:pref}.
\end{definition}

\ber While $\fF_{\omega,t}^\ttm(\alpha)$, $\fF_{\omega_+,t}^\ttm(\alpha)$
and  $\fF_{\omega_+,t}^\anc(\alpha)$ are obviously positive for
$\alpha\in\RR$, the quantity $\fF_{\omega_+,t}^\qpsc(\alpha)$ is a priori
complex. The principal branch of the logarithm should be understood in the
definition of $\bF^\qpsc_{\omega_+}$. This makes
$\log\fF^\qpsc_{\omega_+,t}(\alpha)$ well-defined for $\lambda$ small
enough\footnote{This is not restrictive since all our results will be
perturbative in $\lambda$.} and $t$ large, see~\eqref{eq:ness-qpsc} and
Remark~\ref{rem:subseq}.
\eer

Definition~\ref{def-PREF} has several aspects. By the Gärtner-Ellis
theorem, the existence and differentiability of the first limit
in~\eqref{def:functionals-limit} give that the family of measures
$\left(Q_{\omega,t}^\ttm(t\,\cdot\;)\right)_{t>0}\subset\cP(\RR)$ satisfies a
large deviation principle on the interval $]a,b[$, where $]a,b[\, =\RR$ if
$I=\RR$, and otherwise
\[
a=\lim_{\alpha\downarrow\vartheta_-}\partial_\alpha\bF_\omega^\ttm(\alpha), \qquad b=\lim_{\alpha\uparrow\vartheta_+}\partial_\alpha\bF_\omega^\ttm(\alpha),
\]
with the rate function
\[
 \II(s)=\sup_{-\alpha\in I}(s\alpha-\bF_\omega^\ttm(-\alpha)).
\]
This is the quantum Evans--Searles fluctuation theorem. When the system is TRI
one moreover has $\fF_{\omega,t}^\ttm(\alpha)=\fF_{\omega,t}^\ttm(1-\alpha)$
for all real $\alpha$ and all $t$, see~\cite[Theorem~1.4]{Benoist2023a}. This
leads to the celebrated symmetry
\beq
{\mathbb I}(-s)=s+ {\mathbb I}(s),
\label{ES-sym-rel}
\eeq
called the quantum Evans--Searles fluctuation relation, that holds for
$|s|<\min\{-a,b\}$, see~\cite[Proposition~2.6]{Benoist2024}.

The existence and differentiability of the second  limit
in~\eqref{def:functionals-limit} give that
$\left(Q_{\omega_+,t}^\ttm(t\,\cdot\;)\right)_{t>0}\subset\cP(\RR)$ satisfies a
large deviation principle on the interval $]a_+,b_+[$, where $]a_+,b_+[\, =\RR$
if $I=\RR$, and otherwise
\[
a_+=\lim_{\alpha\downarrow\vartheta_-}\partial_\alpha\bF_{\omega_+}^\ttm(\alpha),\qquad b_+=\lim_{\alpha\uparrow\vartheta_+}\partial_\alpha\bF_{\omega_+}^\ttm(\alpha),
\]
with the rate function
\[
\II_+(s)=\sup_{-\alpha\in I} (s\alpha-\bF^\ttm_{\omega_+}(-\alpha)).
\]
This is the quantum Gallavotti--Cohen fluctuation theorem.
Theorem~\ref{prelim-thm}(3) identifies these two fluctuation  theorems:
$\bF^\ttm_{\omega_+}=\bF^\ttm_{\omega}$, $a=a_+$, $b= b_+$, $\II=\II_+$. If the
system is TRI, the symmetry $\II_+(-s)=s+\II_+(s)$ therefore also holds. This is
the quantum Gallavotti--Cohen fluctuation relation.

The other equalities in~\eqref{def:pref} link  the~2TMEP with~EAST and~QPSC. For
a more thorough discussion about the PREF we refer the reader
to~\cite{Benoist2024} (see in particular Section~2.8).

\subsection{Main results}
\label{ssec:main-results}

We start by introducing our final assumptions. The first two are linked to the
complex spectral deformation of Liouvilleans in the so-called ``glued''
Araki-Wyss GNS representation of $\cO_j$ induced by $\omega_j$ and originally
introduced in \cite{Jaksic1996b, Jaksic1996a, Jaksic2002a}. The third assumption
is the Fermi golden rule condition which ensures that the small system $\sS$ is
effectively coupled to the reservoirs. The fourth and last assumption will
ensure time-reversal invariance of the coupled system when needed.

\assuming{SFM5}{SFMfive}{There exists a Hilbert space $\fH$ such that, for
$1\leq j\leq M$, $\fh_j=L^2(\RR_+,\d s)\otimes\fH$ and $h_j$ is the
operator of multiplication by the variable $s\in\RR_+$.
}

The assumption that  $\fH_j=\fH$ for all $j$ is made only for notational
convenience. The other parts of~\SFMfive{} will play an essential role in our
analysis. In what follows we will often write $\fh$ for $\fh_j$ and $h$ for
$h_j$. Note that~\SFMfive{} implies~\SFMfour.

We assume that $\cK_\sS$ and $\fH$ are equipped with complex conjugations which
we denote by $\overline{\,\cdot\vphantom{X}\;}$. These anti-linear involutions
extend naturally to $\cB(\cK_\sS)$, $\fh$, $\Gamma_-(\fh)$ and their
tensor products. We will also denote by $\overline{\,\cdot\vphantom{X}\;}$ these
extensions.\footnote{The assumption that this conjugation is the same for all
reservoirs is only for notational convenience, and one could choose a different
conjugation in each reservoir.} To each $f\in\fh$ we associate the function
$\widetilde{f}\in L^2(\RR,\d s)\otimes\fH$ defined as
\beq
\widetilde{f}(s)=
\begin{cases}
f(s)&\text{if }s\geq0,\\[1mm]
\BAR{f(|s|)}&\text{if }s<0.
\end{cases}
\label{tildef}
\eeq
Let $\fI(r)=\{z\in\CC\mid |\Im z|<r\}$ and denote by $H^2(r)$ the Hardy class
of all analytic functions $f:\fI(r)\to\fH$ such that
\[
\|f\|^2_{H^2(r)}=\sup_{|\theta|<r}\int_\RR\|f(s+\i\theta)\|_{\fH}^2\d s<\infty.
\]

Our next regularity assumption is

\assuming{SFM6}{SFMsix}{For all $r>0$ and all $j,k,m$, $\widetilde{f}_{j,k,m}\in H^2(r)$.
In addition, for all $r>0$ and all $a>0$,
\[
\sup_{|\theta|<r}\int_\RR\e^{a|s|}\|\widetilde{f}_{j,k,m}(s+\i\theta)\|_\fH^2\d s<\infty.
\]
}

Note in particular that~\SFMsix{} implies~\SFMthree, and hence~\SFMtwo.

\medskip
We now turn to the Fermi golden rule condition. Invoking~\SFMzero, the fermionic Wick theorem gives
\beq\label{eq:wick}
\omega_j\left(R_{j,k}\tau_j^t(R_{j,l})\right) =\delta_{kl}\prod_{m=1}^{n_{j,k}}\omega_j\left(\varphi_j\left(f_{j,k,m}\right)\varphi_j\left(\e^{\i th_j}f_{j,k,m}\right)\right).
\eeq
In Section~\ref{sssec:quasienergy-davies} we shall see that Assumptions~\SFMfive--\SFMsix{}
imply that, for any $0\le a<\pi/\beta_j$,
$$
\cC_{j,k,m}(t)=\omega_j\left(\varphi_j\left(f_{j,k,m}\right)\varphi_j\left(\e^{\i th_j}f_{j,k,m}\right)\right)=O\left(\e^{-a|t|}\right),
$$
as $|t|\to\infty$. Note also, see e.g. \cite{Davies1974,Jaksic2014a}, that
$$
c_{j,k}(u) =\int_\RR\e^{-\i ut}\omega_j\left(R_{j,k}\tau_j^t(R_{j,k})\right)\, \d t\ge0
$$
for all $u\in\RR$. Our Fermi golden rule assumption is

\assuming{SFM7}{SFMseven}{
\begin{enumerate}[{\rm (a)}]
\item $c_{j,k}(u)>0$ for all $u\in\{E'-E\mid E,E'\in\sp(H_\sS)\}$ and all $j,k$.%
\footnote{$\sp(A)$ denotes the spectrum of the operator $A$.}
\item For all $j\in\{1,\ldots,M\}$,%
\footnote{$\cA^\prime$ denotes the commutant of the set $\cA\subset\cB(\cK_\sS)$.}
\[
\{Q_{j,k}\mid 1\leq k\leq m_j\}^\prime\cap\{H_\sS\}^\prime =\CC\one.
\]
\end{enumerate}
}

\ber\SFMseven(a) is usually formulated in terms of the non-negative matrices
$h_j(u)=[h_j^{(kl)}(u)]$ where
\[
h_j^{(kl)}(u)=\int_\RR\e^{-\i ut}\omega_j(R_{j,k}\tau_j^t(R_{j,l}))\d t,
\]
and requires $h_j(u)>0$ for all $u\in\{E'-E\mid E,E'\in\sp(H_\sS)\}$
and all $j$. \SFMzero{} however implies that $h_j(u)$ is diagonal,
see~\eqref{eq:wick}, hence $h_j(u)>0$ indeed reduces to $c_{j,k}(u)>0$ for all $k$.
\eer

\medskip
Note that the spin--fermion model may not be time-reversal invariant.
The next assumption ensures it is.

\assuming{SFM8}{SFMeight}{
The complex conjugation of $\fH$ is such that $\BAR{f}_{j,k,m}=f_{j,k,m}$
for all $j,k,m$. Moreover, the complex conjugation on $\cK_\sS$ is such
that $H_\sS$ and $\i^{n_{j,k}(n_{j,k}-1)/2}Q_{j,k}$ are real with respect
to the induced complex conjugation on $\cB(\cK_\sS)$.
}

Our main result is
\begin{theorem}\label{main-thm-sfm}
Suppose that~\SFMfive--\SFMseven{} hold. Then:
\ben
\item There exists $\Lambda >0$ such that \SFMone{} holds for $0<|\lambda|<\Lambda$.

\item For any $\vartheta >0$ there exists $\Lambda >0$ such that  the PREF
holds on $]-\vartheta, 1+\vartheta[$ for $0 <|\lambda|<\Lambda$. Moreover,
the function
\[
]-\vartheta, 1+\vartheta[\, \ni \alpha \mapsto  \bF(\alpha)
\]
is real-analytic. It is identically vanishing on $]-\vartheta, 1+\vartheta[$
if $\beta_1=\cdots=\beta_M$, and otherwise strictly convex on this interval.

\item If the system is TRI, in particular if~\SFMeight{} holds, it moreover
satisfies the symmetry
$$
\bF(\alpha)=\bF(1-\alpha).
$$
\een
\end{theorem}

Part~(1) was established in~\cite{Jaksic2002a} and is stated here for
completeness. We will prove Part~(2) using the techniques developed
in~\cite{Jaksic1996b,Jaksic1996a,Jaksic2002a} and following the axiomatic
quantum transfer operators resonance scheme of~\cite{Benoist2024}; see also
Section~\ref{ssec:remarks}. As already mentioned, Part~(3) follows readily from
time-reversal invariance. It is an equivalent formulation of~\eqref{ES-sym-rel},
and for this reason is also often called fluctuation relation.

\ber\label{rem:follow}
\ben
\item It follows from~\eqref{def:ttm-measure} that the function $\bF$ is
real-valued and convex on $]-\vartheta,1+\vartheta[$. It satisfies
$\bF(0)=\bF(1)=0$ and, due to convexity, $\bF(\alpha)\leq 0$ for
$\alpha\in[0,1]$ and $\bF(\alpha)\geq0$ for $\alpha\not\in[0,1]$. The fact that
$\bF$ is either identically vanishing or is strictly convex on $]-\vartheta,
1+\vartheta[$ then follows from its analyticity. Finally, $\bF$ is strictly
convex iff $\omega_+(\sigma)>0$, see Section \ref{ssec:nonvanishing}.

\item Although the interval $]-\vartheta,1+\vartheta[$ on which the PREF holds
can be taken arbitrarily large, our result is not uniform in the sense that
$\Lambda$ has to be taken smaller and smaller as $\theta$ grows. This
restriction resembles the high temperature one that is present
in~\cite{Jaksic2002a}.

\item For a discussion of the dependence of $\Lambda$ on the $\beta_j$'s see
\cite{Jaksic2002a,Jaksic2006a}.
\een
\eer

\medskip
\noindent{\bf Example: The simplest spin--fermion model}

In its simplest version the spin--fermion model $\sS$ is a two-level system,
{\sl i.e.,} $\cK_\sS=\CC^2$, with Hamiltonian $H_\sS=\sigma_z$. The interaction
between $\sS$ and each reservoir $\sR_j$ is of the form
\[
V_j = \sigma_x \otimes \varphi_j(f_j),
\]
{\sl i.e.,} in~\eqref{sf-form}--\eqref{sf-form-1} and for all $j,k$ we have
$m_j=n_{j,k}=1$, $Q_{j,k}=\sigma_x$, and we assume that the form factors $f_j$
satisfy~\SFMsix{} and \SFMeight. The operators $H_\sS$ and $Q_{j,k}=\sigma_x$
are real with respect to the usual conjugation on $M_2(\CC)$ so that the system
is TRI. Finally, \SFMseven{} reduces to $\|\tilde f_j(2)\|^2_{\fH}>0$ for all
$j$.

Under these assumptions Theorem~\ref{main-thm-sfm} holds true. Although it does
not appear in our notation, $\bF$ also depends on $\lambda$, and we have
\[
\bF(\alpha)=\lambda^2\bF_2(\alpha)+O(\lambda^3)
\]
where
\beq\label{eq:simplesf-functional}
\begin{split}
\bF_2(\alpha)=-\frac{\pi}2\left(\vphantom{\sqrt{\sum_{j,k=1}^M}} \sum_{j=1}^M\right.&\|\tilde f_j(2)\|^2_{\fH_j}\\
&\left.-\sqrt{\sum_{j,k=1}^M \left(\tanh(\beta_j)\tanh (\beta_k) +\frac{\cosh((\beta_j-\beta_k)(1-2\alpha))}{\cosh(\beta_j)\cosh(\beta_k)}\right) \|\tilde f_j(2)\|^2_{\fH}\|\tilde f_k(2)\|^2_{\fH}}\,\right).
\end{split}
\eeq

\section{Quantum transfer operators approach to the PREF}
\label{sec:qto}

As mentioned at the end of the previous section, the proof of
Theorem~\ref{main-thm-sfm} is based on the study of complex resonances of a
suitable family of non self-adjoint operators called $\alpha$-Liouvilleans.
These operators are generators of one-parameter groups of so-called quantum
transfer operators~\cite{Benoist2024}. The $\alpha$-Liouvilleans are defined on
the GNS Hilbert space $\cH$ and are a generalization of the C-Liou\-vil\-lean
introduced in~\cite{Jaksic2002a} in the study of NESS. In this section we
briefly recall the ``glued'' Araki--Wyss GNS representation of the free Fermi
gas, introduce the $\alpha$-Liouvilleans of~\cite{Benoist2024} in the context of
the spin--fermion model, and recall their connection with the entropic
functionals.

\subsection{The ``Glued'' Araki--Wyss representation}
\label{ssec:gns-glued}

The original Araki--Wyss representation was introduced in~\cite{Araki1964a}. For
pedagogical introductions to the topic we refer
to~\cite{Bratteli1981,Bratteli1987,Jaksic2010b, Derezinski2013}. The ``glued''
form of this representation was introduced in~\cite{Jaksic2002a} and is an
essential step in our spectral approach.

Let $\fh=L^2(\RR_+,\d s)\otimes\fH$, $h$ be the operator of multiplication by
the variable $s\in\RR_+$ (recall~\SFMfive), and let $\omega$ be the quasi-free
state on $\CAR(\fh)$ generated by
\[
T=\left(\one +\e^{\beta h}\right)^{-1}
\]
where $\beta>0$. The $C^\ast$-dynamics $\tau$ is the group of Bogoliubov
$\ast$-automorphisms induced by $h$. We recall that $\omega$ is the unique
$(\tau, \beta)$-KMS state on ${\rm CAR}(\fh)$.

Setting $\widetilde{\fh}=L^2(\RR,\d s)\otimes\fH$, to any $f\in\fh$ we associate
the pair $(f_{\beta},f_{\beta}^\#)\in\widetilde{\fh}\times\widetilde{\fh}$ given
by
\beq
f_\beta(s)=\left(\e^{-\beta s}+1\right)^{-1/2}\widetilde{f}(s),\qquad
f_\beta^\#(s)=\i\left(\e^{\beta s}+1\right)^{-1/2}\widetilde{f}(s),
\label{tildebeta}
\eeq
where $\tilde{f}$ is defined in~\eqref{tildef}. Note that
$f_\beta^\#(s)=\i\overline{ f_\beta(-s)}$. The ``glued'' Araki--Wyss
representation of $\CAR(\fh)$ induced by $\omega$ is the triple
$(\cH,\pi,\Omega)$, where $\cH=\Gamma_-(\widetilde{\fh})$ is the antisymmetric
Fock space over $\widetilde{\fh}$, $\Omega\in\cH$ is the Fock vacuum vector, and
\[
\pi(\varphi(f))=\tphi\left(f_{\beta}\right)=\frac1{\sqrt2}\left(\widetilde a^\ast\left(f_\beta\right)+\widetilde a\left(f_\beta\right)\right),
\]
$\widetilde a^\ast/\widetilde a$ denoting the fermionic creation/annihilation
operator, and $\tphi$ the associated Segal field operator on the Fock space
$\Gamma_-(\tilde\fh)$.

In this representation the standard Liouvillean of $\tau$ is
\[
\cL=\d\Gamma(s),
\]
where, with a slight abuse of notation, $s$ denotes the operator of
multiplication by $s$ on $\widetilde{\fh}$. The modular operator of the state
$\omega$ is
\[
\Delta_\omega=\e^{-\beta\cL}=\Gamma(\e^{-\beta s}),
\]
and the modular group acts as
\[
\varsigma_\omega^\theta (\tphi(f_\beta))
=\Delta_\omega^{\i\theta}\tphi(f_\beta)\Delta_\omega^{-\i\theta}
=\tphi(\e^{-\i\theta\beta s}f_\beta).
\]
Finally, the modular conjugation $J$ is such that
\[
J\tphi(f_\beta)J=\i\e^{\i\pi N}\tphi(f_\beta^\#),
\]
where $N=\d\Gamma(\one)$ is the number operator on $\Gamma_-(\widetilde{\fh})$.

We finish with a remark regarding the thermal factors in~\eqref{tildebeta}.

\ber\label{rem:thfactor}
The maps
\[
\RR \ni s \mapsto \left(\e^{\pm \beta s} +1\right)^{-1/2}
\]
have analytic extension to the strip $|\Im\,z |<\pi/\beta$ and for any $0<r<\pi/\beta$,
\beq
\sup _{|\Im z|\leq r}\left|\left(\e^{\pm \beta z} +1\right)^{-1/2}\right|<\infty.
\label{bs-fact}
\eeq
This basic fact will play an important role in what follows.
\eer

\subsection{The modular structure of the spin--fermion model}
\label{ssec:spin-fermion-qto}

For computational purposes it is convenient to work in the following
GNS-representation $(\cH_\sS,\pi_\sS,\Omega_\sS)$ of the small system algebra
$\cO_\sS=\cB(\cK_\sS)$ associated to the faithful state $\omega_\sS$. The GNS
Hilbert space is $\cH_\sS=\cO_\sS$, equipped with the Hilbert-Schmidt inner
product $\langle X,Y\rangle=\tr(X^\ast Y)$. The representation is the left
multiplication, $\pi_\sS(A)X=AX$, and the cyclic vector is
$\Omega_\sS=\omega_\sS^{1/2}=\dim(\cK_\sS)^{-1/2}\one$. The modular operator of
$\omega_\sS$ is trivial, $\Delta_{\omega_\sS}X=X$, the modular conjugation is
$J_\sS X=X^\ast$ and the standard Liouvillean of $\tau_\sS$ is $\cL_\sS X=[H_\sS,X]$.
Note in particular that $\sp(\cL_\sS)=\{E-E'\mid E,E'\in\sp(H_\sS)\}$.

\medskip For each $1\leq j\leq M$ we denote by $(\cH_j,\pi_j,\Omega_j)$ the
``glued'' Araki--Wyss representation of $\cO_j$ induced by $\omega_j$, as
described in the previous section. We also denote by $\tphi_j$, $\cL_j$,
$\Delta_j$ and $J_j$ the associated field operator, standard Liouvillean,
modular operator and conjugation. The GNS representation of the $C^\ast$-algebra
$\cO$ of the spin--fermion model induced by the reference state $\omega$ is
\[
\begin{split}
\cH&= \cH_{\sS}\otimes \cH_1\otimes \cdots\otimes \cH_{M},\\[1mm]
\pi&=\pi_\sS \otimes \pi_1\otimes \cdots \otimes \pi_M,\\[1mm]
\Omega &=\Omega_\sS \otimes \Omega_1\otimes \cdots \otimes \Omega_M.
\end{split}
\]
We adopt the shorthand $\Omega_\sR=\Omega_1\otimes \cdots \otimes \Omega_M$.
The modular operator and modular conjugation of the state $\omega$ are
\[
\Delta_\omega=\Delta_{\omega_\sS}\otimes\Delta_1\otimes\cdots\otimes\Delta_M,\qquad
J=J_\sS\otimes J_1\otimes\cdots\otimes J_M,
\]
and the modular group acts as
\[
\varsigma_\omega^{\theta}\left(\pi\left(A\otimes\varphi(f_1)
\otimes\cdots\otimes\varphi(f_M)\right)\right)
=\pi_\sS(A)\otimes\tphi_1\big(\e^{-\i\theta\beta_1s}f_{1,\beta_1}\big)
\otimes\cdots\otimes\tphi_M\big(\e^{-\i\theta\beta_Ms}f_{M,\beta_M}\big).
\]

The standard Liouvillean of the free dynamics $\tau_\fr$ is
\[
\cL_\fr=\cL_\sS+\cL_1+\cdots+\cL_M,
\]
and the standard Liouvillean of the interacting dynamics $\tau_\lambda$ is
\[
\cL_\lambda=\cL_\fr+\lambda\pi(V)-\lambda J\pi(V)J.
\]
Note that $\pi(V)$ is a sum of terms of the form
\[
\i ^{n_{j,k}(n_{j,k}-1)/2}\pi_\sS(Q_{j,k})\otimes\tphi_j
\big(f_{j,k, 1, \beta_j}\big)\cdots\tphi_j\big(f_{j,k, n_{j,k}, \beta_j}\big)
\]
corresponding to~\eqref{sf-form}--\eqref{sf-form-1}.
Similarly $J\pi(V)J$ is a sum of terms of the form
\beq
\i^{n_{j,k}(n_{j,k}-1)/2}J_\sS\pi_\sS(Q_{j,k})J_\sS
\otimes\left[\i\e^{\i\pi N_j}\tphi_j\big(f_{j,k,1,\beta_j}^\#\big)\right]
\cdots\left[\i \e^{\i \pi N_j} \tphi_j\big(f_{j,k, n_{j,k}, \beta_j}^\#\big) \right],
\label{correspond-1}
\eeq
where $J_\sS\pi_\sS(A)J_\sS X=XA^\ast$, and $N_j$ is the number operator on
$\cH_j$. We will denote by $\overline{\,\cdot\vphantom{X}}$ the complex
conjugation on $\cH$ naturally associated to the ones on $\cK_\sS$ and $\fH$.

\subsection{Two families of Liouvilleans}
\label{ssec:alpha-liouv}

Central to the proof of Theorem~\ref{main-thm-sfm} is the following family of
$\alpha$-Liouvilleans $\cL_{\lambda,\alpha}$ of~\cite{Benoist2024}, first
introduced in~\cite{Jaksic2010b}. For $\alpha\in\i\RR$ they are given by
$$
\cL_{\lambda,\alpha} =\cL_\fr+\lambda\left(\pi(V)-J\varsigma_\omega^{-\i\bar\alpha}(\pi(V))J\right).
$$
Note that, in analogy with~\eqref{correspond-1},
\beq\label{vsharp}
J\varsigma_\omega^{-\i\bar\alpha}(\pi(V))J
\eeq
is a sum of terms of the form
\beq\label{correspond-2}
\i ^{n_{j,k}(n_{j,k}-1)/2}J_\sS\pi_\sS(Q_{j,k})J_\sS
\otimes\left[\i\e^{\i\pi N_j}\tphi_j\big(\e^{-\alpha\beta_js}f_{j,k,1,\beta_j}^\#\big)\right]
\cdots \left[\i\e^{\i\pi N_j}\tphi_j\big(\e^{-\alpha\beta_js}f_{j,k,n_{j,k},\beta_j}^\#\big)\right].
\eeq
Under Assumption~\SFMsix, $\cL_{\lambda,\alpha}$ is defined for all
$\alpha\in\CC$ by analytic continuation of~\eqref{vsharp} in the variable
$\alpha$. Note that, due to the linearity/anti-linearity of the map
$\widetilde\fh\ni f\mapsto\widetilde a_j^\ast(f)/\widetilde
a_j(f)\in\pi_j(\cO_j)$, the analytic continuation of the factor
$\tphi_j\big(\e^{-\alpha\beta_js}f_{j,k,m,\beta_j}^\#\big)$ in the
product~\eqref{correspond-2} is given, for arbitrary $\alpha\in\CC$, by
\[
\frac1{\sqrt 2}\left(\widetilde a_j\big(\e^{\bar\alpha\beta_js}f_{j,k,m,\beta_j}^\#\big) +\widetilde a_j^\ast\big(\e^{-\alpha\beta_j s}f_{j,k,m,\beta_j}^\#\big)\right).
\]
In what follows, for $\alpha\in\CC$, we set
$$
W_j(\alpha)=\pi(V_j)-J\varsigma_\omega^{-\i\bar\alpha}(\pi(V_j))J,\qquad
W(\alpha)=\sum_{j=1}^MW_j(\alpha).
$$
For arbitrary $\alpha\in\CC$, the $\alpha$-Liouvillean $\cL_{\lambda,\alpha}$
generates a bounded, strongly continuous one-parameter group $\left(\e^{\i
t\cL_{\lambda,\alpha}}\right)_{t\in\RR}$ on $\cH$, unitary for $\alpha\in\i\RR$.
These so-called quantum transfer operators $\e^{\i t\cL_{\lambda,\alpha}}$ will
play a particular role in the analysis of the 2TMEP, EAST and QPSC functionals.

We also introduce the closely related operators
\beq
\widehat\cL_{\lambda,\alpha}
=\Delta_\omega^{-\alpha/2}\cL_{\lambda,1/2-\alpha}\Delta_\omega^{\alpha/2}
=\cL_\fr+\lambda\Delta_\omega^{-\alpha/2}W\left(\tfrac12-\alpha\right)\Delta_\omega^{\alpha/2}.
\label{def:Lalphahat}
\eeq

The primary reason for introducing these $\alpha$-Liouvilleans is the following
representation of the entropic functionals~\cite[Proposition~3.2 and Equations~(5.11)--(5.12)]{Benoist2024}.
\bep\label{prop:liouv-functionals}
For all $\alpha\in\CC$ and $t\in \RR$
\[
[D\omega_{-t}:D\omega]_\alpha\Omega=\e^{\i t\cL_{\lambda,1/2-\alpha}}\Omega,
\qquad
[D\omega_{-t}:D\omega]^\ast_{\frac{\bar\alpha}{2}}[D\omega_{-t}:D\omega]_{\frac{\alpha}{2}}\Omega
=\e^{\i t\widehat\cL_{\lambda,\alpha}}\Omega.
\]
As a consequence, for all $\alpha\in\CC$ and $t,T\in\RR$,
\ben
\item $\ds\fF_{\omega,t}^\ttm(\alpha)=\langle\Omega,\e^{\i t\cL_{\lambda,1/2-\alpha}}\Omega\rangle$.
\item $\ds\fF_{\omega_T,t}^\qpsc(\alpha)=\langle\Omega,\e^{\i T\cL_{\lambda,1/2}}[D\omega_{-t}:D\omega]_\alpha\Omega\rangle =\langle\Omega,\e^{\i T\cL_{\lambda,1/2}}\e^{\i t\cL_{\lambda,1/2-\alpha}}\Omega\rangle$.
\item $\ds\fF_{\omega_T,t}^\anc(\alpha)=\langle\Omega,\e^{\i T\cL_{\lambda,1/2}} [D\omega_{-t}:D\omega]^\ast_{\frac{\bar\alpha}{2}}[D\omega_{-t}:D\omega]_{\frac{\alpha}{2}}\Omega\rangle
=\langle\Omega,\e^{\i T\cL_{\lambda,1/2}}\e^{\i t\widehat\cL_{\lambda,\alpha}}\Omega\rangle$.
\een
\eep

\ber All the above identities are first derived for $\alpha\in\i\RR$, and then
extended to $\alpha\in\CC$ by analytic continuation. Indeed, as already
mentioned, \SFMthree{} implies~\cite[Assumption~\AnV]{Benoist2024} for arbitrary
$\vartheta>0$, which guarantees that all the involved quantities have entire
analytic extensions.
\eer

\section{Liouvillieans: spectra, resonances and dynamics}
\label{sec:deformedliouvillean}

The proof of our main Theorem~\ref{main-thm-sfm} relies on the representation of
the entropic functionals given in Proposition~\ref{prop:liouv-functionals}, on
complex deformations of the $\alpha$-Liouvilleans, and on the analysis of the
spectral resonances unveiled by these deformations. The general strategy is
adapted from~\cite{Jaksic1996b,Jaksic1996a,Jaksic2002a}, to which we refer for
more details.

For the reader's convenience, in this section we briefly recall the construction
of the deformed Liouvilleans and their main properties, introducing their
spectral resonances. The central results of this section concern the large time
behaviour of the associated quantum transfer operators and are given in Sections
\ref{ssec:liouv-dyn}-\ref{ssec:alphahat-liouv}.

\subsection{Complex deformation of $\cL_{\lambda,\alpha}$}
\label{ssec:complex-deformation}

Following on Remark~\ref{rem:thfactor}, we set $\hat r=\min_j(\pi/\beta_j)$ and
denote by $\cF$ the collection of all the $f_{j,k,m,\beta_j}$ and
$f_{j,k,m,\beta_j}^\#$. It then follows from~\SFMsix{} and~\eqref{bs-fact} that,
for any $0<r<\hat r$ and $a>0$, $\cF\subset H^2(r)$ and
\[
\sup_{\twol{f\in\cF}{|\theta|\leq r}}\int_\RR\e^{a|s|}\|f(s+\i\theta)\|_\fH^2\d s<\infty.
\]
Let $p=\i\partial_s$ be the  generator of the translation group $(\e^{-\i\theta
p}f)(s)=f(s+\theta)$. Set $P=\d\Gamma(p)$ and define the unitary group
$(U(\theta))_{\theta\in\RR}$ on $\cH$ by $U(\theta)=\one\otimes\e^{-\i\theta
P}\otimes\cdots\otimes\e^{-\i\theta P}$. Further setting
\beq
W(\alpha,\theta)=U(\theta)W(\alpha)U(-\theta)
\label{eq:Walphatheta},
\eeq
we observe that
\begin{align*}
\cL_\fr(\theta)& =U(\theta)\cL_\fr U(-\theta)=\cL_\fr+\theta N,\\[2mm]
\cL_{\lambda,\alpha}(\theta)
&=U(\theta)\cL_{\lambda,\alpha}U(-\theta)
=\cL_\fr+\theta N+\lambda W(\alpha,\theta),
\end{align*}
where $N=\sum_j N_j$. The map~\eqref{eq:Walphatheta} has an analytic extension
\beq
\CC\times\fI(\hat r)\ni(\alpha,\theta)\mapsto W(\alpha,\theta)\in\cB(\cH),
\eeq
which is bounded on $B\times\fI(r)$, for any $0<r<\hat r$ and any bounded
$B\subset\CC$. This allows us to define $\cL_{\lambda,\alpha}(\theta)$ for
$\lambda,\alpha\in\CC$ and $\theta \in \fI(\hat r)$. In what follows, we write
\[
\fI^+(r)=\{z\in\CC\mid0<\Im z<r\}.
\]

We now summarize the basic properties of the family $\left(\cL_{\lambda,
\alpha}(\theta)\right)_{\theta\in\fI(\hat r)}$ of complex deformations of the
$\alpha$-Liouvillean $\cL_{\lambda,\alpha}$.

\paragraph{\bf(a)} For $\Im\theta\neq0$, $\cL_\fr(\theta)=\cL_\fr+\theta N$ is a
closed normal operator with domain $\Dom(\cL_\fr)\cap\Dom(N)$. Its discrete
spectrum is $\sp(\cL_\sS)$ and its essential spectrum is the union of
lines $\RR+\i\,\Im(\theta)\NN^\ast$.\footnote{$\NN^\ast=\NN\setminus\{0\}.$}
Moreover, $\cL_\fr(\theta)^\ast=\cL_\fr(\bar\theta)$, and for
$z\not\in\sp(\cL_\fr(\theta))$, one has
\beq\label{est-nor}
\|(z-\cL_\fr(\theta))^{-1}\|=\frac1{\dist(z,\sp(\cL_\fr(\theta)))}.
\eeq

\paragraph{\bf(b)} For $(\lambda,\alpha,\theta)\in\CC\times\CC\times\fI^+(\hat
r)$, $\cL_{\lambda,\alpha}(\theta)$ is a closed operator with domain
$\Dom(\cL_\fr)\cap\Dom(N)$, and adjoint
$\cL_{\lambda,\alpha}(\theta)^\ast=\cL_{\bar\lambda,\bar\alpha}(\bar\theta)$.
Moreover, with
$$
C_{\alpha,\theta}=\|W(\alpha,\theta)\|,
$$
one has
$$
\sp(\cL_{\lambda,\alpha}(\theta))\subset D_{\lambda,\alpha,\theta} =\{z\in\CC\mid\dist(z,\sp(\cL_\fr(\theta)))\leq|\lambda|C_{\alpha,\theta}\},
$$
and for $z\in\CC\setminus D_{\lambda,\alpha,\theta}$,
\[
\|(z-\cL_{\lambda,\alpha}(\theta))^{-1}\|\leq\frac1{\dist(z,\sp(\cL_\fr(\theta)))-|\lambda|C_{\alpha, \theta}}.
\]
All these results follow from~{\bf(a)} and standard estimates based
on the resolvent identity.

\paragraph{\bf(c)} For $(\lambda,\alpha,\theta)\in\CC\times\CC\times\fI^+(\hat r)$,
$\cL_{\lambda, \alpha}(\theta)$ is an analytic family of type $A$ in each
variable separately (see, e.g.~\cite[Section~XII.2]{Reed1978}).

\medskip\noindent
In the following we fix $0<r<\hat r$ as well as $\vartheta,\zeta>0$, and set
\beq
C=\sup_{\twol{|\Im \theta|\leq r}{\alpha\in B(\vartheta,\zeta)}} C_{\alpha, \theta},
\qquad
B(\vartheta,\zeta)=\{z\in\CC\mid|\Re z|<\vartheta, |\Im z|<\zeta\}.
\label{def-C}
\eeq

\paragraph{\bf(d)} Suppose that $\alpha\in B(\vartheta,\zeta)$.
Then, if\, $\Im z<-|\lambda|C$,
\[
\slim_{\Im\theta\downarrow0}\,(z-\cL_{\lambda,\alpha}(\theta))^{-1}=(z-\cL_{\lambda,\alpha}(\Re\theta))^{-1}.
\]
The proof is the same as that of~\cite[Lemma~4.8]{Jaksic1996b}.

\begin{figure}
\begin{center}
\includegraphics[width=1\textwidth]{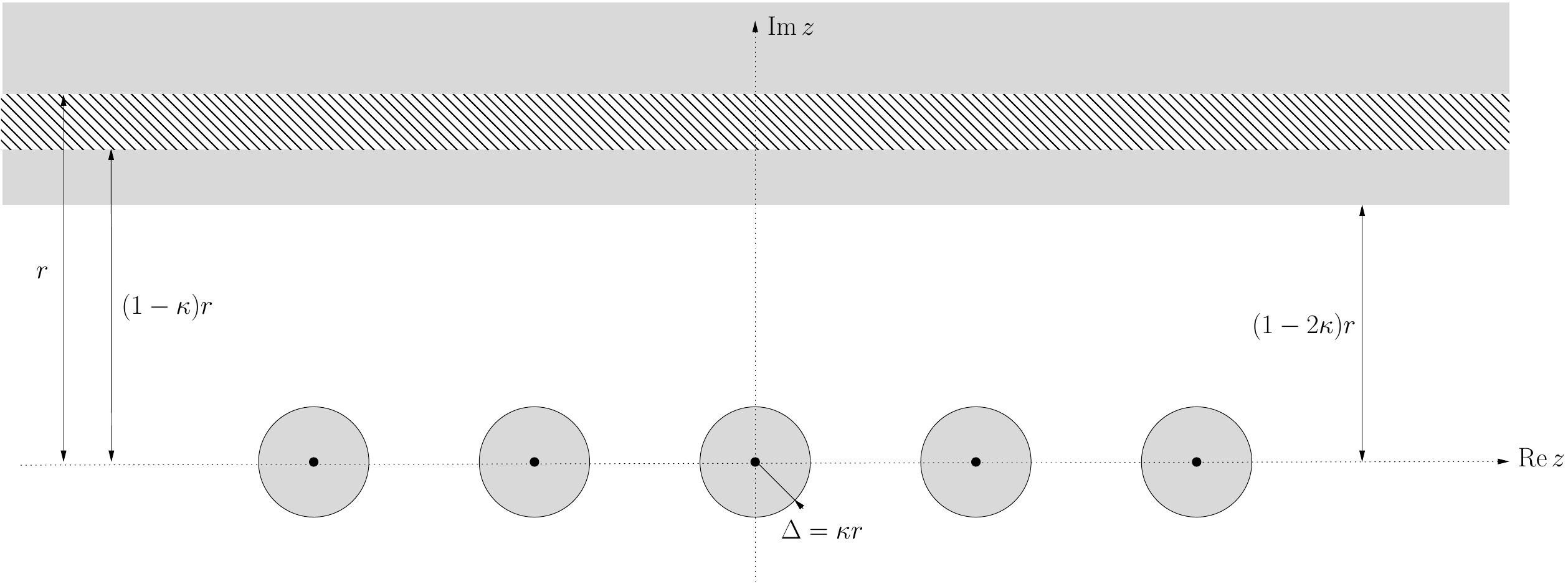}
\end{center}
\caption{Picture of the $z$-plane (it is assumed here that $\kappa r<r_\sS/4$).
The black dots are the eigenvalues of $\cL_\sS$. If $\theta$ is in the hashed
area, the spectrum of the deformed Liouvillean $\cL_{\lambda,\alpha}(\theta)$ is
in the shaded areas (which extends to infinity on the top side).
}
\label{Fig:Spec}
\end{figure}

\paragraph{\bf(e)}\label{parae}\ASS{parae}{e} Let
\[
r_\sS=\min\left\{|e-e'|\,\bigg|\, e,e'\in\sp(\cL_\sS),e\neq e'\right\},
\qquad
0<\kappa<\frac16,
\]
and $\Lambda>0$ be such that
\[
\Lambda C=\Delta=\min\left\{\kappa r,\frac{r_\sS}{4}\right\}.
\]
It then follows from~{\bf(b)} that for $|\lambda|<\Lambda$,
$\alpha\in B(\vartheta,\zeta)$, and $(1-\kappa)r<\Im\theta\leq r$,
one has
\beq
\sp(\cL_{\lambda,\alpha}(\theta))\subset \left\{w\in\CC\mid\Im w>(1-2\kappa)r\right\}\cup\left\{w\in\CC\mid\dist(w,\sp(\cL_\sS))<\Delta\right\}.
\label{eq:SpecSeparation}
\eeq
Moreover, for all
\[
z\in\left\{w\in\CC\mid\Im w\leq(1-3\kappa)r,\; \dist(w,\sp(\cL_\sS))\geq r_\sS/2\right\},
\]
and for all $|\lambda|<\Lambda$, $\alpha\in B(\vartheta,\zeta)$,
$(1-\kappa)r<\Im\theta\leq r$, the estimate
\beq
\|(z-\cL_{\lambda,\alpha}(\theta))^{-1}\|\leq\frac1{\Delta},
\label{est-res-v}
\eeq
holds, see Figure~\ref{Fig:Spec}.

The first condition $\Delta\le\kappa r$ ensures that the two subsets on the
right-hand side of~\eqref{eq:SpecSeparation} are disjoint. It follows that the
spectrum of $\cL_{\lambda,\alpha}(\theta)$ in the subset
$\{z\in\CC\mid\dist(z,\sp(\cL_\sS))<\Delta\}$ is discrete. The second condition
$\Delta\le r_\sS/4$ ensures that for any distinct $e,e'\in\sp(\cL_\sS)$
$$
\{z\in\CC\mid|z-e|<\Delta\}\cap\{z\in\CC\mid|z-e'|<\Delta\}=\emptyset,
$$
so that the spectral projection
$$
\cQ_{\lambda,\alpha,e}(\theta)
=\oint\limits_{|z-e|=r_\sS/2}(z-\cL_{\lambda,\alpha}(\theta))^{-1}\frac{\d z}{2\pi\i}
$$
onto the part of spectrum of $\cL_{\lambda,\alpha}(\theta)$ inside the disk
$|z-e|<r_\sS/2$ has exactly the same rank as the spectral projection
$\one_{e}(\cL_\sS)$ of $\cL_\sS$ for the eigenvalue $e$. This spectrum actually
coincides with the spectrum of a linear map
\beq\label{eq:quasienergy}
\Sigma_e(\lambda,\alpha): \ran\one_e(\cL_\sS)\to\ran\one_e(\cL_\sS),
\eeq
called quasi-energy operator in~\cite{Jaksic2002a}, that does not depend on
$\theta$. Hence, the spectrum of $\cL_{\lambda,\alpha}(\theta)$ in the
half-plane $\Im z<(1-2\kappa)r$ is discrete and independent of $\theta$ as long
as $(1-\kappa)r<\Im\theta\leq r$. The finitely many eigenvalues in this
half-plane are called {\sl spectral resonances} of $\cL_{\lambda,\alpha}$. We
shall briefly recall the construction of the quasi-energy operators
$\Sigma_e(\lambda,\alpha)$ in Section~\ref{sssec:quasienergy-def}.

\paragraph{\bf(f)}\label{paraf}\ASS{paraf}{f} We start with the observation
that, for $\Im\theta\geq0$, $\left(\e^{\i t\cL_\fr(\theta)}\right)_{t\geq0}$ is
a  strongly continuous contraction semi-group on $\cH$. It then follows that for
all $(\lambda,\alpha)\in\CC^2$ and $0\leq\Im\theta<\hat r$, $\left(\e^{\i
t\cL_{\lambda,\alpha}(\theta)}\right)_{t\geq0}$ is also a strongly continuous
semi-group on $\cH$. For $t,\lambda,\theta\in\RR$ and $\alpha\in\i\RR$, one has
$$
\e^{\i t\cL_{\lambda,\alpha}(\theta)}
=\fG_{\lambda,\alpha,\theta}^t\e^{\i t\cL_\fr(\theta)}
$$
with the unitary cocycle
$$
\fG_{\lambda,\alpha,\theta}^t
=\one+\sum_{n\ge1}(\i\lambda t)^n\int\limits_{0\le s_1\le\cdots\le s_n\le 1}
W_{ts_1}(\alpha,\theta)\cdots W_{ts_n}(\alpha,\theta)\d s_1\cdots\d s_n,
$$
where
$$
W_t(\alpha,\theta)=\e^{\i t\cL_\fr(\theta)}W(\alpha,\theta)\e^{-\i t\cL_\fr(\theta)}.
$$
By Assumptions~\SFMfive--\SFMsix, the map $(\alpha,\theta)\mapsto
W_t(\alpha,\theta)\in\cB(\cH)$ extends to an analytic function on $\CC\times
\fI(\hat r)$, bounded for $(\alpha,\Im\theta)$ in any compact subset of
$\CC\times\, ]-\hat r,\hat r[$. It follows that for $t\in\RR$,
$(\lambda,\alpha,\theta)\mapsto\fG_{\lambda,\alpha,\theta}^t$ is an analytic
function on $\CC^2\times \fI(\hat r)$, bounded for $(\lambda,\alpha,\Im\theta)$
in any compact subset of $\CC^2\times \, ]-\hat r,\hat r[$. Note that since
$U(\theta')W_t(\alpha,\theta)U(-\theta')=W_t(\alpha,\theta+\theta')$ for
$\theta'\in\RR$, one has
$$
U(\theta')\fG^t_{\lambda,\alpha,\theta}\Psi
=\fG^t_{\lambda,\alpha,\theta+\theta'}U(\theta')\Psi.
$$
Thus, if $\Psi$ is analytic for the group $(U(\theta))_{\theta\in\RR}$
in a strip $0\le\Im\theta<\rho\leq\hat r$, so is
$\fG^t_{\lambda,\alpha,\theta}\Psi$, and the latter identity
extends by analyticity.

The family $\left(\e^{\i\theta N}\right)_{\Im\theta\ge0}$ is a strongly
continuous contraction semi-group which is analytic in the open upper half-plane
$\Im\theta>0$. Thus, if $\Psi\in\cH$ is analytic for the group
$(U(\theta))_{\theta\in\RR}$ in a strip $0\le\Im\theta<\rho$, then, for $t>0$,
the map
$$
\RR\ni\theta\mapsto\e^{\i t\cL_\fr(\theta)}U(\theta)\Psi
=\e^{\i t\theta N}\e^{\i t\cL_\fr}U(\theta)\Psi
$$
has a bounded continuous extension to the strip $0\le\Im\theta <\rho$
which is analytic in its interior. It follows that the same holds for the map
$$
\RR\ni\theta\mapsto U(\theta)\e^{\i t\cL_\fr}\Psi,
$$
and hence that the respective extensions satisfy
$$
\e^{\i t\cL_\fr(\theta)}U(\theta)\Psi=U(\theta)\e^{\i t\cL_\fr}\Psi
$$
for $0\le\Im\theta<\rho$. Combined with the previous results, we conclude that
$$
\fG_{\lambda,\alpha,\theta}^t\e^{\i t\cL_\fr(\theta)}U(\theta)\Psi
=\fG_{\lambda,\alpha,\theta}^tU(\theta)\e^{\i t\cL_\fr}\Psi
=U(\theta)\fG_{\lambda,\alpha,0}^t\e^{\i t\cL_\fr}\Psi,
$$
and so
\beq
\e^{\i t\cL_{\lambda,\alpha}(\theta)}U(\theta)\Psi
=U(\theta)\e^{\i t\cL_{\lambda,\alpha}}\Psi.
\label{eq:passthru}
\eeq

For later use, see in particular Section~\ref{ssec:proof-qpsc}, we summarize
the above discussion in the following lemma.
\bel\label{lem:semigroup-anlytic-continuation}
If\, $\Psi$ is analytic for the group $(U(\theta))_{\theta\in\RR}$ in a strip
$0\leq \Im\theta<\rho\leq \hat r$, then the map
\[
\RR \ni \theta \mapsto U(\theta)\e^{\i t \cL_{\lambda, \alpha}}\Psi\in \cH
\]
has an analytic extension to the same strip, which is bounded and continuous
on any closed substrip, and for any $\theta$ in this strip
\[
\e^{\i t \cL_{\lambda, \alpha}(\theta)} U(\theta)\Psi
=U(\theta)\e^{\i t \cL_{\lambda, \alpha}}\Psi.
\]
\eel

\subsection{The quasi-energy operators}
\label{ssec:quasienergy}

In this section we first recall briefly the construction of the maps
$\Sigma_e(\lambda, \alpha)$. They go back to~\cite{Hunziker1983} and we refer
the reader to e.g.~\cite{Hunziker1983,Jaksic1996b,Jaksic2002a} for more details.
In a second part we study their connection with the Davies generator and the
level-shift operator, see
e.g.~\cite{Davies1974,Derezinski2003,Derezinski2006,Jaksic2014a}. This
connection plays a key role in the study of spectral properties of
$\Sigma(\lambda,\alpha)$, hence of $\cL_{\lambda,\alpha}(\theta)$, and which is
given in Section~\ref{sssec:quasienergy-spectrum}.

The standing assumptions in this section are the ones made in Paragraph~\parae{}
of the previous section.

\subsubsection{Construction of $\Sigma_e(\lambda, \alpha)$}
\label{sssec:quasienergy-def}

The map $(\lambda,\alpha)\mapsto\cQ_{\lambda,\alpha,e}(\theta)$ is analytic and
\[
\cQ_{0,\alpha,e}(\theta)=\cQ_{0,0,e}(\theta)
=\one_{e}(\cL_\fr)=\one_e(\cL_\sS)\otimes|\Omega_\sR\rangle\langle\Omega_\sR|.
\]
It follows from the estimate~\eqref{est-res-v} and the resolvent identity
that, by possibly making $\Lambda$ smaller,
\[
\sup_{\twol{|\lambda|<\Lambda}{\alpha\in B(\vartheta,\zeta)}}
\|\cQ_{\lambda,\alpha,e}(\theta)-\cQ_{0, 0, e}(\theta)\|<1
\]
for all $e\in\sp(\cL_\sS)$. This gives that the map
\[
S_{\lambda,\alpha,e}(\theta)
=\cQ_{0,0,e}(\theta)\cQ_{\lambda,\alpha,e}(\theta):
\ran\cQ_{\lambda,\alpha,e}(\theta)\to\ran\cQ_{0,0,e}(\theta)
\]
is an isomorphism, reducing to the identity for $\lambda=0$.
The {\sl quasi-energy} operator~\eqref{eq:quasienergy} is defined by
\beq
\Sigma_e(\lambda,\alpha)=S_{\lambda,\alpha,e}(\theta)\cQ_{\lambda,\alpha,e}(\theta) \cL_{\lambda,\alpha}(\theta)\cQ_{\lambda,\alpha,e}(\theta)S_{\lambda,\alpha,e}(\theta)^{-1}.
\label{def:quasienergy}
\eeq
As the notation suggests, $\Sigma_e(\lambda,\alpha)$ does not depend on
$\theta$, see e.g.~\cite{Jaksic1996b}. In the following, it will be convenient
to identify $\ran S_{\lambda,\alpha,e}(\theta)
=\ran\one_e(\cL_\fr)=\ran\one_e(\cL_\sS)\otimes\Omega_\sR$ with
$\ran\one_e(\cL_\sS)$, so that $\Sigma_e(\lambda,\alpha)$ will act on the
eigenspace of $\cL_\sS$ for its eigenvalue $e$.

By construction these quasi-energy operators satisfy
$$
\sp\big(\Sigma_e(\lambda,\alpha)\big)
=\sp\left(\cL_{\lambda,\alpha}(\theta)\right)\cap\big\{z\in\CC\,
\big|\,|z-e|<r_\sS/2\big\},
$$
and have the following properties which follow from regular
perturbation theory~\cite{Kato1966, Reed1978,Hunziker1983}.
\ben
\item The map $(\lambda,\alpha)\mapsto\Sigma_e(\lambda,\alpha)$ is analytic and
\beq
\Sigma_e(\lambda,\alpha)=e\one_e(\cL_\sS)+\lambda^2 \Sigma_{e}^{(2)}(\alpha)+O(\lambda^3),
\label{eq:quasienergy-taylor}
\eeq
where the estimate $O(\lambda^3)$ is uniform in $\alpha\in B(\vartheta,\zeta)$.
The term linear in $\lambda$ vanishes due to Assumption~\SFMzero, and it is at
this point that~\SFMzero{} enters critically in the proof.

\item A Fermi golden rule computation gives
\beq
\Sigma_{e}^{(2)}(\alpha) =\lim_{\epsilon\uparrow0}\one_e(\cL_\sS)T_\sS^\ast W(\alpha) (e+\i\epsilon-\cL_\fr)^{-1}W(\alpha)T_\sS\one_e(\cL_\sS),
\label{eq:lso}
\eeq
where $T_\sS:\cH_\sS\ni X\mapsto X\otimes\Omega_\sR$ has the adjoint $T_\sS^\ast X\otimes\Psi_\sR=\langle\Omega_\sR,\Psi_\sR\rangle X$.

\item
$$
\Sigma(\lambda,\alpha)=\bigoplus_{e\in\sp(\cL_\sS)}\Sigma_e(\lambda,\alpha),\qquad \Sigma^{(2)}(\alpha)=\bigoplus_{e\in\sp(\cL_\sS)}\Sigma_e^{(2)}(\alpha),
$$
defines two operators acting on $\cH_\sS$. The operator $\Sigma^{(2)}(\alpha)$
is the so-called {\sl level-shift} operator for the triple
$(\cH_\sS,\cL_\fr,W(\alpha))$, see e.g.~\cite{Derezinski2006}.
Equation~\eqref{eq:lso} also allows to define $\Sigma_e^{(2)}(\alpha)$, hence
$\Sigma^{(2)}(\alpha)$, for all $\alpha\in\CC$.
\een

\subsubsection{Quasi-energy operators and deformed Davies generators}
\label{sssec:quasienergy-davies}

In this section we turn to the close relation between $\Sigma^{(2)}(\alpha)$ and
the deformed Davies generator introduced in~\cite{Jaksic2014a}. We will use this
connection in the next section to study spectral properties of
$\Sigma^{(2)}(\alpha)$. Those of $\Sigma(\lambda,\alpha)$ will then follow from
regular perturbation theory.

It follows from~\SFMfive{} and~\eqref{tildef} that
\begin{align*}
\cC_{j,k,m}(t)
&=\omega_j\left(\varphi_j(f_{j,k,m})\varphi_j(\e^{\i th_j}f_{j,k,m})\right)\\[2pt]
&=\tfrac12\langle f_{j,k,m},(\e^{\i th_j}(1-T_j)+\e^{-\i th_j}T_j)f_{j,k,m}\rangle\\[2pt]
&=\tfrac14\int_\RR\frac{\cosh\left((\beta_j/2+\i t)s\right)}{\cosh\left(\beta_j s/2 \right)} \tilde{f}_{j,k,m}(-s)\tilde{f}_{j,k,m}(s)\d s\\[2pt]
&=\tfrac12\int_\RR\frac{\tilde{f}_{j,k,m}(-s)\tilde{f}_{j,k,m}(s)}{1+\e^{-\beta_js}}\e^{\i ts}\d s,
\end{align*}
and~\SFMsix{} allows to deform the integration contour from $\RR$ to
$\RR\pm\i a$, as long as $|a|<\pi/\beta_j$,
$$
\cC_{j,k,m}(t)=\tfrac12\int_\RR\frac{\tilde{f}_{j,k,m}(-(s\pm\i a))\tilde{f}_{j,k,m}(s\pm\i a)}
{1+\e^{-\beta_j(s\pm\i a)}}\e^{\i t(s\pm\i a)}\d s.
$$
This gives that $\cC_{j,k,m}(t)=O(\e^{-a|t|})$ for all $j,k,m$, provided
$0\le a<\hat r$. Hence, it holds that

\assuming{SFM00}{SFMzerozero}{
For some $\epsilon>0$ and all $j,k,m$,
$$
\int_0^\infty|\cC_{j,k,m}(t)| t^\epsilon\d t<\infty.
$$
}

Denote by $\tau_{j,\lambda}$ the $C^\ast$-dynamics on $\cO_\sS\otimes\cO_j$
generated by $\delta_\sS+\delta_j+\i\lambda[V_j,\,\cdot\;]$.
Assumption~\SFMzero{} and~\SFMzerozero{} go back to~\cite{Davies1974}, where it
was shown that, for all $X,Y\in\cO_\sS$,
\beq
\lim_{\twol{\lambda\to0}{\lambda^2t=\xi}}
\omega_\sS\otimes\omega_j\left( X\tau_{0,j}^{-t}\circ\tau_{\lambda,j}^t(Y)\right)
=\omega_\sS\left(X\e^{\xi K_j}(Y)\right),
\label{dav-vanh-1}
\eeq
for some $K_j\in\cB(\cO_\sS)$. By similar arguments, see~\cite{Derezinski2006},
one can show that
$$
\lim_{\twol{\lambda\to0}{\lambda^2t=\xi}}\omega\left(X \tau_0^{-t}\circ\tau_{\lambda}^t(Y)\right)
=\omega_\sS\left(X\e^{\xi K}(Y)\right),
$$
with
$$
K=\sum_jK_j.
$$
$K_j$ is the {\sl Davies generator} of the system $\sS+\sR_j$ and $K$ that of
the full system $\sS+\sR$. The semi-groups $\left(\e^{t K_j}\right)_{t\geq0}$
and $\left(\e^{t K}\right)_{t\geq0}$ are completely positive and unital on
$\cO_\sS$. If Assumption~\SFMseven{} holds, these semi-groups are also
positivity improving, see e.g.~\cite{Spohn77b,Jaksic2014a}.

For $\alpha\in\CC$, following~\cite{Jaksic2014a}, we define the deformed Davies
generators $K_{\alpha,j}$ and $K_{\alpha}$ acting on $\cO_\sS$ by
\beq
K_{\alpha,j}(X)=K_j\left(X\e^{\alpha\beta_jH_\sS}\right)\e^{-\alpha\beta_jH_\sS},
\qquad
K_\alpha=\sum_jK_{\alpha,j}.
\label{def:deformed-davies}
\eeq
We note that $K_j$ commutes with $\delta_\sS$ (see~\cite[Theorem~2.1]{Davies1974}
or~\cite[Theorem~6.1]{Derezinski2006}) so that
\beq
K_{\alpha+z,j}(X)=\e^{-z\beta_jH_\sS}K_{\alpha,j}\left(\e^{z\beta_jH_\sS}X\right),
\label{eq:gauge-invariant-davies}
\eeq
for all $z\in\CC$. For $\alpha\in\RR$, the semi-groups $\left(\e^{t
K_{\alpha,j}}\right)_{t\geq0}$ and $\left(\e^{t K_\alpha}\right)_{t\geq0}$ are
also completely positive, and they are moreover positivity improving if
Assumption~\SFMseven{} holds, see~\cite[Theorem~3.1]{Jaksic2014a}.

The following proposition gives the announced connection between these deformed
Davies generators and the level-shift operators.

\bep\label{prop:lso-davies}
Suppose that~\SFMzero, \SFMfive{} and~\SFMsix{} hold. Then, for all $\alpha\in\RR$,
\beq
\Sigma^{(2)}(\alpha)=-\i K_{1/2-\alpha}.
\label{eq:lso-davies}
\eeq
\eep

\begin{corollary}\label{cor:lso-cp}
Suppose that~\SFMzero, \SFMfive{} and~\SFMsix{} hold. Then, for all
$\alpha\in\RR$ the semi-group $\left(\e^{\i t\Sigma^{(2)}(\alpha)}\right)_{t\geq0}$
is completely positive, and unital when $\alpha=\frac12$. If Assumption~\SFMseven{}
holds, then this semi-group is also positivity improving.
\end{corollary}

The relation~\eqref{eq:lso-davies} can be proven by direct computation, see
e.g.~\cite[Section~6.7]{Derezinski2003} or~\cite{Derezinski2006}. Alternatively,
one can give a structural proof following~\cite{Derezinski2004}
where~\eqref{eq:lso-davies} was established in the cases $\alpha=0$ and
$\alpha=1/2$.\footnote{Note that our convention for the Davies generator differs
from the one in~\cite{Derezinski2004} by a factor $\i$.} For the reader's
convenience we finish this section with a proof of
Proposition~\ref{prop:lso-davies} along these lines.

\ber Proposition~\ref{prop:lso-davies} can actually be proven under much more
general condition than~\SFMsix. However, making this assumption  does not affect
the generality of our main result while  allowing  for a relatively simple proof
of Lemma~\ref{lem:deformed-davies} below which is the main technical ingredient
of the argument.
\eer

Recall that $\Omega_j$ is the vector representative of the state $\omega_j$ in
the GNS Hilbert space of the $j^\mathrm{th}$ reservoir. Setting
\[
\cL_{\lambda,\alpha,j}=\cL_\sS+\cL_j+\lambda W_j(\alpha),
\]
one has the following lemma, compare with~\eqref{dav-vanh-1}.

\begin{lemma}\label{lem:deformed-davies}
For all $X,Y\in\cB(\cK_\sS)$ and $\alpha\in\i\RR$,
$$
\lim_{\twol{\lambda\to0}{\lambda^2t=\xi}}\langle X\otimes\Omega_j, \e^{-\i t\cL_\sS}\e^{\i t\cL_{\lambda,\alpha,j}}Y\otimes\Omega_j\rangle =\tr(X^\ast\e^{\i\xi\Sigma^{(2)}_j(\alpha)}Y),
$$
where $\Sigma_j^{(2)}(\alpha)$ is the level-shift operator for $(\cH_\sS,\cL_\sS+\cL_j,W_j(\alpha))$,
\beq
\Sigma_j^{(2)}(\alpha)=\bigoplus_{e\in\sp(\cL_\sS)}\lim_{\epsilon\uparrow 0}\one_e(\cL_\sS) T_\sS^\ast W_j(\alpha)(e+\i\epsilon-(\cL_\sS+\cL_j))^{-1}W_j(\alpha)T_\sS\one_e(\cL_\sS).
\label{eq:SigmajDef}
\eeq
\end{lemma}

\proof The proof is an application of~\cite[Theorem~3.4]{Derezinski2006}. In
that theorem the only assumption which requires a comment is the following one:
for some $\lambda_0>0$
\[
\int_0^\infty\sup_{|\lambda|\leq\lambda_0} \left\|\cP W_j(\alpha)\e^{\i t(1-\cP)\cL_{\lambda,\alpha,j} (1-\cP)}W_j(\alpha)\cP\right\| \, \d t<\infty,
\]
where $\cP=I_\sS\otimes|\Omega_j\rangle\langle\Omega_j|$. To check it, we introduce
complex deformations as in Section~\ref{ssec:complex-deformation},
\begin{align*}
\cL_{\lambda,\alpha,j}(\theta)&=U(\theta)\cL_{\lambda,\alpha,j}U(-\theta)
=\cL_{0,j}(\theta)+\lambda W_j(\alpha,\theta),\notag\\[2mm]
\cL_{0,j}(\theta)&=\cL_\sS+\cL_j+\theta N_j.
\end{align*}
For $\Im\theta\geq0$, $\left(\e^{\i t\cL_{0,j}(\theta)}\right)_{t\geq0}$ is a
strongly continuous contraction semi-group on $\cH_\sS\otimes\cH_j$. Hence, for
all $\lambda,\alpha\in\CC$ and $0\leq\Im\theta<\hat r$, the perturbed family
$\left(\e^{\i t\cL_{\lambda,\alpha,j}(\theta)}\right)_{t\geq0}$ is a strongly
continuous semi-group on $\cH_\sS\otimes\cH_j$. Moreover, since $\cP
U(\theta)=\cP=U(-\theta)\cP$, we can write
\beq
\cP W_j(\alpha)\e^{\i t(1-\cP)\cL_{\lambda,\alpha,j}(1-\cP)}W_j(\alpha)\cP
=\cP W_j(\alpha,\theta)\e^{\i t(1-\cP)\cL_{\lambda,\alpha,j}(\theta)(1-\cP)}W_j(\alpha,\theta)\cP.
\eeq
It follows from~\SFMzero{} that $\cP V_j\cP=0$,\footnote{That $\cP V_j\cP=0$ is
also necessary to get~\eqref{dav-vanh-1}, see~\cite{Davies1974}.} and similarly
$\cP W(\alpha,\theta)\cP=0$. It therefore suffices to prove that
\beq
\int_0^\infty\sup_{|\lambda|\leq\lambda_0}\left\|
(1-\cP)\e^{\i t(1-\cP)\cL_{\lambda,\alpha,j}(\theta)(1-\cP)}(1-\cP)
\right\|\,\d t<\infty,
\label{eq:proof-df}
\eeq
for some $\lambda_0>0$ and some $\theta$ such that $0\leq\Im\theta<\hat r$.
Observing that the operator appearing in the last formula belongs to the
semi-group acting on $\ker\cP$ and generated by
$$
(1-\cP)\cL_{\lambda,\alpha,j}(\theta)|_{\ker\cP}
=\cL_\sS+\d\Gamma(s+\theta)|_{\Omega_j^\perp}
+\lambda(1-\cP)W(\alpha,\theta)|_{\ker\cP},
$$
we get, for $\lambda=0$ and $t\ge0$,
\[
\left\|(1-\cP)\e^{\i t(1-\cP)\cL_{0,j}(\theta)(1-\cP)}(1-\cP)\right\|
=\e^{-t\Im(\theta)},
\]
and a standard estimate on perturbed semi-groups~\cite[Theorem~3.1]{Davies1980} gives
\[
\|(1-\cP)\e^{\i t(1-\cP)\cL_{\lambda,\alpha,j}(\theta)(1-\cP)}(1-\cP)\|
\leq\e^{-t(\Im\theta-|\lambda|\|W(\alpha,\theta)\|)},
\]
so that the estimate~\eqref{eq:proof-df} follows by fixing $\theta$ such
that $0<\Im(\theta)<\hat r$ and $0<\lambda_0$ small enough.
\qed

\medskip
\noindent{\bf Proof of Proposition~\ref{prop:lso-davies}.}
For $\alpha\in\i\RR$ we have that
\[
V_j^\#(\alpha)
=(J_\sS\otimes J_j)(\e^{\alpha\beta_j\cL_j}V_j\e^{-\alpha\beta_j\cL_j})(J_\sS\otimes J_j)
=\e^{\alpha\beta_j\cL_j}(J_\sS\otimes J_j )V_j(J_\sS\otimes J_j)\e^{-\alpha\beta_j\cL_j}.
\]
Consider now the operator $H_\sS+\cL_j+\lambda V_j$ where, with the usual abuse
of notation, $H_\sS$ stands for $\pi(H_\sS)$, and so
$H_\sS+\lambda V_j\in\pi(\cO)$. By the Trotter product formula,
\begin{align*}
\e^{\alpha\beta_j(H_\sS+\cL_j+\lambda V_j)}(J_\sS\otimes J_j)
&V_j(J_\sS\otimes J_j)\e^{-\alpha\beta_j(H_\sS+\cL_j+V_j)}\\[1mm]
=\lim_{n\to\infty}\left(\e^{\alpha\beta_j(H_\sS+\lambda V_j)/n}
\e^{\alpha\beta_j\cL_j/n}\right)^n (J_\sS\otimes J_j )
&V_j(J_\sS\otimes J_j) \left(\e^{-\alpha\beta_j(H_\sS+V_j)/n}
\e^{-\alpha\beta_j\cL_j/n}\right)^n.
\end{align*}
Since, for $t\in\RR$, $\e^{\i t\cL_j}\pi(\cO)'\e^{-\i t\cL_j}\subset\pi(\cO)'$,
we derive that for $\alpha\in\i\RR$,
\beq
V_j^\#(\alpha)=\e^{\alpha\beta_j(H_\sS+\cL_j+\lambda V_j)}(J_\sS\otimes J_j) V_j(J_\sS\otimes J_j)\e^{-\alpha\beta_j(H_\sS+\cL_j+\lambda V_j)}.
\label{n-nev}
\eeq
Using that $\cL_\sS+\cL_j+\lambda V_j=\cL_{\lambda,\alpha,j}-\lambda V_j^\#(\alpha)$
commutes with $H_\sS+\cL_j+\lambda V_j$, \eqref{n-nev} further yields that
\[
\cL_{\lambda,\alpha,j}
=\e^{\alpha\beta_j(H_\sS+\cL_j+\lambda V_j)}\cL_{\lambda,0,j} \e^{-\alpha\beta_j(H_\sS+\cL_j+\lambda V_j)}.
\]
Combined with Lemma~\ref{lem:deformed-davies}, this relation gives that,
for $\alpha\in\i(-\zeta,\zeta)$,
\beq
\Sigma_j^{(2)}(\alpha)(X)
=\e^{\alpha\beta_jH_\sS}\Sigma_j^{(2)}(0)(\e^{-\alpha\beta_jH_\sS}X).
\label{eq:lso-similarity}
\eeq
By analyticity, this relation holds for all $\alpha\in\RR$, provided its left-hand side
is given by~\eqref{eq:SigmajDef}. It was shown in~\cite[Theorem~3.1]{Derezinski2004} that
$$
\Sigma_j^{(2)}(0)=-\i K_{1/2,j},
$$
and so
$$
\Sigma^{(2)}(\alpha)(X)=\sum_j\Sigma_j^{(2)}(\alpha)(X)
=\sum_j\e^{\alpha\beta_jH_\sS}\Sigma_j^{(2)}(0)(\e^{-\alpha\beta_jH_\sS}X)
=-\i\sum_j\e^{\alpha\beta_jH_\sS}K_{1/2,j}(\e^{-\alpha\beta_jH_\sS}X).
$$
Taking into account~\eqref{def:deformed-davies}--\eqref{eq:gauge-invariant-davies},
we finally get
$$
\Sigma^{(2)}(\alpha)=-\i\sum_jK_{1/2-\alpha,j}=-\i K_{1/2-\alpha}.
$$
\qed

\subsubsection{Spectral analysis of $\Sigma(\lambda, \alpha)$}
\label{sssec:quasienergy-spectrum}

As a direct consequence of Corollary~\ref{cor:lso-cp} we first get the following
spectral result about the level-shift operator $\Sigma^{(2)}(\alpha)$, see
e.g.~\cite[Theorem~2.2]{Jaksic2014a}. Let
$$
\cE^{(2)}(\alpha)=\i\,\min\left\{\Im w\mid w\in\sp\left(\Sigma^{(2)}(\alpha)\right)\right\}.
$$

\bel\label{lem:lsospectrum}
Assuming~\SFMzero{} and~\SFMfive--\SFMseven{}, the following assertions hold for $\alpha\in\RR$.
\ben
\item $\cE^{(2)}(\alpha)$ is a purely imaginary simple eigenvalue of\, $\Sigma^{(2)}(\alpha)$,
with $\cE^{(2)}\left(\tfrac12\right)=0$.
\item All the other eigenvalues\,
$z\in\sp\left(\Sigma^{(2)}(\alpha)\right)\setminus\left\{\cE^{(2)}(\alpha)\right\}$
satisfy\, $\Im\left(z-\cE^{(2)}(\alpha)\right)>0$.
\item The eigenprojection for the eigenvalue $\cE^{(2)}(\alpha)$ writes
$P_\alpha=|X_\alpha\rangle\langle Y_\alpha|$, where
$X_\alpha,Y_\alpha\in \cH_\sS=\cB(\cK_\sS)$ are positive definite.
\een
\eel

\bep\label{prop:quasienergyspectrum}
Under the assumptions of the previous lemma, for any $\vartheta,\zeta>0$ there
exists $\Lambda,\epsilon>0$ such that:
\ben
\item For $0<|\lambda|<\Lambda$ and $\alpha\in B(\vartheta,\zeta)$, the linear
map $\Sigma(\lambda,\alpha)$ has a simple eigenvalue $\cE(\lambda,\alpha)$
such that for any other eigenvalue
$w\in\sp(\Sigma(\lambda,\alpha))\setminus\{\cE(\lambda,\alpha)\}$ one has
\[
\Im(w-\cE(\lambda,\alpha))\geq\lambda^2\epsilon.
\]
\item For fixed $\lambda$, the map $B(\vartheta,\zeta)\ni\alpha\mapsto\cE(\lambda,\alpha)$
is analytic.
\item $\cE(\lambda,\alpha)\in\sp(\Sigma_0(\lambda, \alpha))$, in particular
$|\cE(\lambda,\alpha)|<\frac{r_\sS}{4}$.
\een
\eep

\proof (1)--(2) Follow immediately from~\eqref{eq:quasienergy-taylor}, the previous
lemma and regular perturbation theory.

\noindent(3) Since $\cE^{(2)}(\alpha)$ is purely imaginary it must be an eigenvalue
of $\Sigma_0^{(2)}(\alpha)$. Regular perturbation theory ensures that
$\cE(\lambda,\alpha)$ is therefore an eigenvalue of
$\Sigma_0(\lambda,\alpha)$ for $\lambda$ small enough.
\qed

\ber Using~\eqref{eq:quasienergy-taylor} we actually have
$\cE(\lambda,\alpha)=\lambda^2\cE^{(2)}(\alpha)+O(\lambda^3)$.
\eer

\subsection{Dynamics of $\alpha$-Liouvilleans}
\label{ssec:liouv-dyn}

In this and the next two sections, we assume that~\SFMzero{} and~\SFMfive--\SFMseven{}
hold, and we set
$$
D=\bigcap_{|\Im\theta|<\hat r}\Dom U(\theta).
$$
Recall also that $0<r<\hat r$ is fixed and $C$ is given by~\eqref{def-C}.

\bep\label{prop:mero-cont}
For any $\tfrac23 r<\varrho< r$ and $\vartheta,\zeta>0$, there exist constants
$\Lambda,\epsilon>0$ such that, for $\alpha\in B(\vartheta,\zeta)$,
$0<|\lambda|<\Lambda$, and $\Phi,\Psi\in D$, the function
\beq
z\mapsto\langle\Phi,(z-\cL_{\lambda,\alpha})^{-1}\Psi\rangle,
\label{eq:resolvForm}
\eeq
originally analytic for $\Im z<-\Lambda C$, has a meromorphic continuation
to the half-plane $\Im z<\varrho$, and its only possible singularity in the
region $\Im(z-\cE(\lambda,\alpha))<\tfrac12\lambda^2\epsilon$ is a simple
pole at $\cE(\lambda,\alpha)$.
\eep

\proof Fix $0<\kappa<\frac16$ such that $\varrho=(1-2\kappa) r$. Let
$\Lambda,\epsilon$ be as Proposition~\ref{prop:quasienergyspectrum} and
$\Phi,\Psi\in D$. For all $\Im z<-\Lambda C$ and $\theta\in\RR$ one has
$$
\langle\Phi,(z-\cL_{\lambda,\alpha})^{-1}\Psi\rangle =\langle U(\theta)\Phi,(z-\cL_{\lambda,\alpha}(\theta))^{-1}U(\theta)\Psi\rangle.
$$
Note that the functions $\RR\ni\theta\mapsto\Psi_\theta=U(\theta)\Psi$ and
$\RR\ni\theta\mapsto\BAR{\Phi}_\theta=\BAR{U(\theta)\Phi}=U(-\theta)\BAR\Phi$
both have analytic continuations to the strip $\fI(\hat r)$. Thus, the identity
\beq\label{eq:deformedresolvent}
\langle\Phi,(z-\cL_{\lambda,\alpha})^{-1}\Psi\rangle
=\left\langle\BAR{\BAR{\Phi}_\theta},(z-\cL_{\lambda,\alpha}(\theta))^{-1}\Psi_\theta\right\rangle
\eeq
holds for all $\theta\in\fI^+(\hat r)$, $|\lambda|<\Lambda$,
$\alpha\in B(\vartheta,\zeta)$ and $\Im z<-\Lambda C$. Using the results from
Paragraph~\parae{} in Section~\ref{ssec:complex-deformation}, by setting
$\theta=\i r$ the right-hand side of this identity provides a meromorphic
continuation of its left-hand side to the half-plane $\Im z<\varrho$.
Proposition~\ref{prop:quasienergyspectrum} then yields the last assertion.
\qed

\ber The residue of the function~\eqref{eq:resolvForm} at $\cE(\lambda,\alpha)$ is given by
\beq\label{res-v+}
c_{\lambda,\alpha}=\langle U(-\i r)\Phi,\cQ_{\lambda,\alpha}(\i r)U(\i r)\Psi\rangle,
\eeq
where $\cQ_{\lambda,\alpha}(\i r)$ is the spectral projection of
$\cL_{\lambda,\alpha}(\i r)$ for the eigenvalue $\cE(\lambda,\alpha)$.
\eer

\bep\label{prop:resolventbound}
For any $\tfrac23 r<\varrho< r$ and $\vartheta,\zeta>0$, there exist a constants
$\Lambda>0$ such that, for $\alpha\in B(\vartheta,\zeta)$, $0<|\lambda|<\Lambda$,
and $\phi,\psi\in \cH$, the function
$f(z)=\langle\phi,(z-\cL_{\lambda, \alpha}(\i r))^{-1}\psi\rangle$ satisfies%
\footnote{$\partial$ denotes the Wirtinger derivative w.r.t.\;$z$.}
\beq\label{anc-1}
\sup_{\twol{y<\varrho}{j\in\{0,1\}}}\int_{|x|>R}|(\partial^jf)(x+\i y)|^{2-j}\,\d x <+\infty,
\eeq
where $R=1+\|\cL_\sS\|$. As a consequence, for any $\Phi,\Psi\in D$, the function
$g(z)=\langle\Phi,(z-\cL_{\lambda, \alpha})^{-1}\Psi\rangle$ satisfies
\beq\label{anc-2}
\sup_{\twol{y<\varrho}{j\in\{0,1\}}}\int_{|x|>R}|(\partial^jg)(x+\i y)|^{2-j}\,\d x <+\infty.
\eeq
\eep

\proof As in the proof of Proposition~\ref{prop:mero-cont}, we fix $0<\kappa<1/6$
such that $\varrho=(1-2\kappa) r$.

Since $\cL_\fr(\i r)$ is a normal operator, the spectral theorem gives that for $z$
in the resolvent set of $\cL_\fr(\i r)$,
$$
\|(z-\cL_\fr(\i r))^{-1}\psi\|^2
=\int_{\sp(\cL_\fr(\i r))}\frac{\d\mu_{\psi}(\xi)}{|z-\xi|^2},
$$
where $\mu_\psi$ denotes the spectral measure of $\cL_\fr(\i r)$ for the vector $\psi$.

For any $\xi\in\sp(\cL_\fr(\i r))$ and $y<\varrho$, one has
$$
\int_{|x|>R}\frac{\d x}{|x+\i y-\xi|^2}\le\max\left(\frac{\pi}{2\kappa r},2\right),
$$
and hence
\beq\label{eq:free-resolv-int}
\sup_{y<\varrho}\int_{|x|>R}\|(x+\i y-\cL_\fr(\i r))^{-1}\psi\|^2\d x<\infty.
\eeq
By the same argument, we also have
\beq\label{eq:free-resolv-intbar}
\sup_{y<\varrho}\int_{|x|>R}\|\left((x+\i y-\cL_\fr(\i r))^{-1}\right)^\ast\phi\|^2\d x<\infty.
\eeq
Invoking the identity~\eqref{est-nor}, we deduce that if $\Lambda$ is small enough
then, for $|\lambda|<\Lambda$ and $\alpha\in B(\vartheta,\zeta)$, the operator
\[
G(z,\lambda,\alpha)=\left(I-\lambda(z-\cL_\fr(\i r))^{-1}W(\alpha,\i r)\right)^{-1}
\]
is well defined for $|\Re z|>R$, $\Im z<\varrho$, and satisfies
\beq\label{eq:resolv-unif-bound}
\sup_{\twol{|\Re z|>R}{\Im z<\varrho}}\|G(z,\lambda,\alpha)\|\leq2.
\eeq
It follows from the second resolvent identity that
\beq\label{eq:resolv-full-free}
(z-\cL_{\lambda,\alpha}(\i r))^{-1}=G(z,\lambda,\alpha)(z-\cL_\fr(\i r))^{-1}.
\eeq
Combining~\eqref{eq:free-resolv-int}, \eqref{eq:resolv-unif-bound}
and~\eqref{eq:resolv-full-free} gives that~\eqref{anc-1} holds for $j=0$.

Similarly, the operator
$$
\widetilde{G}(z,\lambda,\alpha)
=\left(I-\lambda W(\alpha,\i r)(z-\cL_\fr(\i r))^{-1}\right)^{-1}
$$
also satisfies
$$
\sup_{\twol{|\Re z|>R}{\Im z<\varrho}}\|\widetilde{G}(z,\lambda,\alpha)\|\leq2,
$$
and is such that
$$
(z-\cL_{\lambda,\alpha}(\i r))^{-1}=(z-\cL_\fr(\i r))^{-1}\widetilde{G}(z,\lambda,\alpha).
$$
We can then write
\[
-\partial(z-\cL_{\lambda,\alpha}(\i r))^{-1}
=(z-\cL_{\lambda,\alpha}(\i r))^{-2}
=(z-\cL_\fr(\i r))^{-1}\widetilde{G}(z,\lambda,\alpha)G(z,\lambda,\alpha)(z-\cL_\fr(\i r))^{-1},
\]
so that
$$
|\partial f(z)|=\left|\partial\langle\phi,(z-\cL_{\lambda,\alpha}(\i r))^{-1}\psi\rangle\right|
\leq4\|\left((z-\cL_\fr(\i r))^{-1}\right)^\ast\phi\|\, \|(z-\cL_\fr(\i r))^{-1}\psi\|.
$$
Combining~\eqref{eq:free-resolv-int}, \eqref{eq:free-resolv-intbar} and the Cauchy-Schwarz
inequality, we obtain~\eqref{anc-1} with $j=1$.

Finally, \eqref{anc-2} follows from~\eqref{eq:deformedresolvent} and~\eqref{anc-1} with
$\theta=\i r$, $\phi=U(-\i r)\Phi$ and $\psi=U(\i r)\Psi$.
\qed

Using Propositions~\ref{prop:mero-cont} and~\ref{prop:resolventbound} together
with~\cite[Proposition~4.1]{Benoist2024} we obtain the following dynamical
result.
\bep\label{prop:liouv-dynamics}
For any $\vartheta,\zeta>0$ there exists $\Lambda,\epsilon>0$ such that
for any $0<|\lambda|<\Lambda$, $\alpha\in B(\vartheta,\zeta)$ and
$\Phi,\Psi\in D$ one has
$$
\langle\Phi,\e^{\i t\cL_{\lambda,\alpha}}\Psi\rangle
=\e^{\i t\cE(\lambda,\alpha)}\left(c_{\lambda,\alpha}+O(\e^{-\lambda^2\epsilon t})\right)
$$
as $t\uparrow\infty$, where $c_{\lambda,\alpha}$ is the residue given by~\eqref{res-v+}.
\eep

The next, closely related, result is proven in an identical way and will be used
in the sequel. Recall that, for all $(\lambda,\alpha)\in\CC^2$ and
$0\leq\Im\theta<\hat r$, $\left(\e^{\i t\cL_{\lambda,\alpha}(\theta)}\right)_{t\geq0}$
is also a strongly continuous semi-group on $\cH$. Combining~\eqref{anc-1} in
Proposition~\ref{prop:resolventbound} with the spectral results about
$\cL_{\lambda,\alpha}(\theta)$ obtained in
Sections~\ref{ssec:complex-deformation}--\ref{ssec:quasienergy}, and
using~\cite[Proposition~4.1]{Benoist2024}, we have the following analogue of
Proposition~\ref{prop:liouv-dynamics}.

\bep\label{prop:deformed-liouv-dynamics}
For any $\vartheta,\zeta>0$, there exist constants $\Lambda,\epsilon>0$ such that, for
all $0<|\lambda|<\Lambda$, $\alpha\in B(\vartheta,\zeta)$ and $\phi,\psi\in\cH$ one has
$$
\langle\phi,\e^{\i t\cL_{\lambda,\alpha}(\i r)}\psi\rangle
=\e^{\i t\cE(\lambda,\alpha)}\left(\langle\phi,\cQ_{\lambda,\alpha}(\i r)\psi\rangle
+O(\e^{-\lambda^2\epsilon t})\right)
$$
as $t\uparrow\infty$.
\eep

\subsection{The $\widehat{\cL}_{\lambda,\alpha}$ Liouvilleans}
\label{ssec:alphahat-liouv}

As mentioned in Section~\ref{ssec:alpha-liouv}, the study of the ancilla part of
the PREF relies on the closely related Liouvilleans
$\widehat\cL_{\lambda,\alpha}$. It is easy to see that the analysis of
$\cL_{\lambda,\alpha}$ presented in the previous sections extends line by line
to $\widehat\cL_{\lambda,\alpha}$ and the associated analytically deformed
$\widehat\cL_{\lambda,\alpha}(\theta)$. We denote by
$\widehat\Sigma^{(2)}(\alpha)$ and $\widehat\Sigma(\lambda,\alpha)$ the
corresponding level-shift and quasi-energy operators.

By the definition~\eqref{def:Lalphahat}, $\widehat\cL_{\lambda,\alpha}$ is
obtained from $\cL_{\lambda,\alpha}$ by replacing $W(\alpha)$ by
$\Delta_\omega^{-\alpha/2}W\left(\tfrac12-\alpha\right)\Delta_\omega^{\alpha/2}$.
Since $\Delta_\omega^{\alpha/2}$ commutes with $\cL_\fr$ and
$\Delta_\omega^{\alpha/2}T_\sS=T_\sS$, the associated level-shift operator
is given by
\begin{align*}
\widehat{\Sigma}_{e}^{(2)}(\alpha)
&=\lim_{\epsilon\uparrow0}\one_e(\cL_\sS)T_\sS^\ast\Delta_\omega^{-\alpha/2}W
\left(\tfrac12-\alpha\right)\Delta_\omega^{\alpha/2} (e+\i\epsilon-\cL_\fr)^{-1}
\Delta_\omega^{-\alpha/2}W\left(\tfrac12-\alpha\right)\Delta_\omega^{\alpha/2}T_\sS\one_e(\cL_\sS)\\[4pt]
&=\lim_{\epsilon\uparrow0}\one_e(\cL_\sS)T_\sS^\ast W\left(\tfrac12-\alpha\right)
(e+\i\epsilon-\cL_\fr)^{-1}
W\left(\tfrac12-\alpha\right)T_\sS\one_e(\cL_\sS)\\[4pt]
& =\Sigma_{e}^{(2)}\left(\tfrac12-\alpha\right),
\end{align*}
for $\alpha\in\i\RR$, hence for all $\alpha\in\CC$ by analyticity. In particular
the conclusions of Corollary~\ref{cor:lso-cp} hold, with unital property when
$\alpha=0$, hence so do those of Lemma~\ref{lem:lsospectrum} and
Proposition~\ref{prop:quasienergyspectrum}, {\sl i.e.,} the operator
$\widehat\Sigma(\lambda,\alpha)$ has a simple eigenvalue
$\widehat\cE(\lambda,\alpha)$ such that for any other eigenvalue $w$ of
$\widehat\Sigma(\lambda, \alpha)$ one has
\[
\Im\left(w-\widehat\cE(\lambda,\alpha)\right)\geq\lambda^2\epsilon.
\]
Arguing as in Section~\ref{ssec:complex-deformation}, Paragraph~\paraf{}, we
derive that, for all $\lambda,\alpha\in\CC$ and $0\leq\Im\theta<\hat r$, the
family $\left(\e^{\i t\widehat{\cL}_{\lambda,\alpha}(\theta)}\right)_{t\geq 0}$
is a strongly continuous semi-group on $\cH$ such that, similarly
with~\eqref{eq:passthru},  one has
\beq\label{eq:semigroup-deformation}
U(\theta)\e^{\i t\widehat{\cL}_{\lambda,\alpha}}\Psi
=\e^{\i t\widehat{\cL}_{\lambda,\alpha}(\theta)}U(\theta)\Psi
\eeq
for all $\Psi\in D$ and $0\leq \Im\theta < \hat r$. Finally, we have the following analogue of
Propositions~\ref{prop:liouv-dynamics} and~\ref{prop:deformed-liouv-dynamics}.

\bep\label{prop:liouv-dynamics2}
For any $\vartheta,\zeta>0$, there exists $\Lambda,\epsilon>0$ such that for all
$0<|\lambda|<\Lambda$, $\alpha\in B(\vartheta,\zeta)$ and all $\Phi,\Psi\in D$ one has
\beq\label{eq:liouv-dynamics2}
\langle\Phi,\e^{\i t\widehat{\cL}_{\lambda,\alpha}}\Psi\rangle
=\e^{\i t\widehat{\cE}(\lambda,\alpha)}
\left(\langle U(-\i r)\Phi,\widehat{\cQ}_{\lambda,\alpha}(\i r)U(\i r)\Psi\rangle
+O(\e^{-\lambda^2\epsilon t})\right)
\eeq
as $t\uparrow\infty$, and where $\widehat{\cQ}_{\lambda,\alpha}(\i r)$ is
the spectral projection of $\widehat{\cL}_{\lambda,\alpha}(\i r)$ for its
eigenvalue $\widehat{\cE}(\lambda,\alpha)$.

Similarly, for all $\phi,\psi\in\cH$ one has
\beq\label{eq:deformed-liouv-dynamics2}
\langle\phi,\e^{\i t\widehat{\cL}_{\lambda,\alpha}(\i r)}\psi\rangle
=\e^{\i t\widehat{\cE}(\lambda,\alpha)}
\left(\langle\phi,\widehat{\cQ}_{\lambda,\alpha}(\i r)\psi\rangle
+O(\e^{-\lambda^2\epsilon t})\right),
\eeq
as $t\uparrow\infty$.
\eep

\section{Proof of Theorem~\ref{main-thm-sfm}}
\label {sec:proof-mainthm}

We prove separately the 2TMEP, QPSC and EAST parts of the PREF. These three
parts are proven respectively in Sections~\ref{ssec:proof-ttm},
\ref{ssec:proof-qpsc} and~\ref{ssec:proof-ancilla}. They all rely on the
representation of the various entropic functionals given in
Proposition~\ref{prop:liouv-functionals} and on
Propositions~\ref{prop:liouv-dynamics}, \ref{prop:deformed-liouv-dynamics}
and~\ref{prop:liouv-dynamics2}.

As a preparation for the proof we establish  some analyticity properties  of the
Connes cocycle. Indeed, for the QPSC and Ancilla parts of the PREF we will first
have to consider the large $T$ limit in
Proposition~\ref{prop:liouv-functionals}(2)--(3) in order to obtain suitable
expressions for $\fF_{\omega_+,t}^\anc$ and $\fF_{\omega_+,t}^\qpsc$. This large
$T$ limit also relies on Proposition~\ref{prop:liouv-dynamics}, applied to
$\e^{iT\cL_{\lambda,1/2}}$. For that purpose one needs to prove that the vectors
\[
[D\omega_{-t}:D\omega]_\alpha\Omega\qquad
\text{and}
\qquad[D\omega_{-t}:D\omega]^\ast_{\frac{\bar\alpha}{2}}[D\omega_{-t}:D\omega]_{\frac{\alpha}{2}}\Omega
\]
belong to the subspace $D$. This will be a consequence of the analyticity properties
of the Connes cocycle we establish in the next section.

\subsection{Analyticity of the Connes cocycle}
\label{ssec:connes-analytic}

The main result in this section is
\bep\label{prop:connes-analytic}
Suppose that~\SFMsix{} holds. Then, for any $t\in\RR$, the map
$$
\RR\times\i\RR\times\RR\ni(\lambda,\alpha,\theta)
\mapsto U(\theta)[D\omega_t:D\omega]_\alpha U(-\theta)\in\cB(\cH)
$$
has an extension to $\CC\times\CC\times\fI(\hat r)$ which is analytic in each
variable separately and is uniformly bounded for $(\lambda,\alpha,\Im\theta)$
in compact subsets of\/ $\CC\times\CC\times\,]-\hat r,\hat r[$.
\eep

\ber The quantity $U(\theta)[D\omega_t:D\omega]_\alpha U(-\theta)$
depends on $\lambda$ through the time evolved state $\omega_t$.
\eer

\proof The proof builds on the results established in~\cite[Section~2]{Benoist2024b}.
For $\alpha\in\i\RR$ recall the identity~\eqref{eq:cocycle-dyson}
\beq\label{eq:cocycle-dyson-bis}
[D\omega_t:D\omega]_\alpha=\one
+\sum_{n=1}^\infty\alpha^n\int\limits_{0\leq u_1\leq\cdots\leq u_n\leq 1}
\varsigma_\omega^{-\i u_1\alpha}(Q_t)\cdots\varsigma_\omega^{-\i u_n\alpha}(Q_t)
\, \d u_1\cdots\d u_n,
\eeq
where
\[
Q_t=\int_0^t\tau_\lambda^{-s}(\sigma)\d s,
\qquad\sigma=\lambda\delta_\omega(V).
\]
Let $(\Gamma_s)_{s\in\RR}$ denote the cocycle associated to the local
perturbation $\lambda V$ of the free dynamics $\tau_\fr$, {\sl i.e.,}
the solution of the Cauchy problem
\[
\partial_s\Gamma_s=\i\lambda\Gamma_s\tau_\fr^s(V),\qquad\Gamma_0=\one.
\]
$\Gamma_s$ is a unitary element of $\cO$ with the norm convergent expansion
\beq\label{eq:gamma-dyson}
\Gamma_s=\one+\sum_{n\ge1}(\i\lambda s)^n\int\limits_{0\le v_1\le\cdots\le v_n\le1} \tau_\fr^{sv_1}(V)\cdots\tau_\fr^{sv_n}(V)\d v_1\cdots\d v_n,
\eeq
and  for $u\in \RR$ we have
\beq\label{entropy-evol}
\varsigma_\omega^u(Q_t)=\int_0^t \varsigma_\omega^u(\Gamma_{-s})\tau_\fr^{-s}(\varsigma_\omega^u(\sigma))
\varsigma_\omega^u(\Gamma_{-s}^\ast)\, \d s,
\eeq
see~\cite[Equations~(2.9) and~(2.14)]{Benoist2024b}.

Now, observe that~\SFMsix{} gives that the maps
\begin{eqnarray}
\RR^3\ni(s,\alpha,\theta)&\mapsto&
U(\theta)\tau_\fr^s\circ\varsigma_\omega^\alpha(V)U(-\theta)\in\cB(\cH),\label{eq:analytic-V}\\[4pt]
\RR^3\ni(s,\alpha,\theta)&\mapsto&
U(\theta)\tau_\fr^s\circ\varsigma_\omega^\alpha(\delta_\omega(V))U(-\theta)\in\cB(\cH),\label{eq:analytic-deltaV}
\end{eqnarray}
have extensions to $\CC^2\times\fI(\hat r)$ which are analytic in each variable
separately and are uniformly bounded for $(s,\alpha,\Im\theta)$ in  compact
subsets of $\CC^2\times\, ]-\hat r,\hat r[$. Using
Equation~\eqref{eq:gamma-dyson} we get that, for $(\lambda,u,\theta)\in\RR^3$,
\begin{eqnarray*}
\lefteqn{U(\theta)\varsigma_\omega^u(\Gamma_{-s})U(-\theta)=}\\[2mm]
& & \one+\sum_{n=1}^\infty(-\i\lambda s)^n\int\limits_{0\leq v_1\leq\cdots\leq v_n\leq1}
[U(\theta)\tau_\fr^{-sv_1}\circ\varsigma_\omega^u(V)U(-\theta)]\cdots
[U(\theta)\tau_\fr^{-sv_n}\circ\varsigma_\omega^u(V)U(-\theta)]\d v_1\cdots\d v_n,
\end{eqnarray*}
and it follows from~\eqref{eq:analytic-V} that the map
\[
\RR^3\ni(\lambda,u,\theta)\mapsto U(\theta)\varsigma_\omega^u(\Gamma_{-s})U(-\theta)
\]
has an extension to $\CC^2\times\fI(\hat r)$ which is analytic in each variable
separately and is uniformly bounded for $(\lambda,u,s,\Im\theta)$ in compact
subsets of $\CC^2\times \RR\times\, ]-\hat r,\hat r[$. The same holds for if
$\Gamma_{-s}$ is replaced by its inverse $\Gamma_{-s}^\ast$.
From~\eqref{entropy-evol} we infer that
\[
U(\theta)\varsigma_\omega^u(Q_t)U(-\theta)= \int_0^t[U(\theta)\varsigma_\omega^u(\Gamma_{-s})U(-\theta)]
[U(\theta)\tau_\fr^{-s}\circ\varsigma_\omega^u(\sigma)U(-\theta)]
[U(\theta)\varsigma_\omega^u(\Gamma_{-s}^\ast)U(-\theta)]\, \d s,
\]
and hence deduce, using~\eqref{eq:analytic-deltaV}, that
\[
\RR^3\ni(\lambda,u,\theta)\mapsto U(\theta)\varsigma_\omega^u(Q_t)U(-\theta)
\]
has an extension to $\CC^2\times\fI(\hat r)$ which is analytic in each variable
separately and is uniformly bounded for $(\lambda,u,t,\Im\theta)$ in compact
subsets of $\CC^2\times\RR\times\, ]-\hat r,\hat r[$. Finally, going back
to~\eqref{eq:cocycle-dyson-bis} we have
\begin{eqnarray*}
\lefteqn{U(\theta)[D\omega_t:D\omega]_\alpha U(-\theta)=}\\[2mm]
&&\one+\sum_{n=1}^\infty\alpha^n\int\limits_{0\leq u_1\leq\cdots\leq u_n\leq1}
[U(\theta)\varsigma_\omega^{-\i u_1\alpha}(Q_t)U(-\theta)]\cdots
[U(\theta)\varsigma_\omega^{-\i u_n\alpha}(Q_t)U(-\theta)]\d u_1\cdots\d u_n,
\end{eqnarray*}
and Proposition~\ref{prop:connes-analytic} follows.
\qed

\medskip
Since $U(-\theta)\Omega=\Omega$, the following consequence of the
previous proposition is immediate.

\begin{corollary}\label{coro:analytic-vectors}
The vectors $[D\omega_{-t}:D\omega]_\alpha\Omega$ and
$[D\omega_{-t}:D\omega]^\ast_{\frac{\bar\alpha}2}[D\omega_{-t}:D\omega]_{\frac{\alpha}2}\Omega$
belong to the subspace $D$, for all $t\in\RR$ and $\alpha\in\CC$.
\end{corollary}

\subsection{2TMEP part of the PREF}
\label{ssec:proof-ttm}

In this and the next two sections we fix $0<r<\hat r$ as in
Section~\ref{sec:deformedliouvillean}.

The starting point is the representation given in
Proposition~\ref{prop:liouv-functionals}(1). Given $\vartheta,\zeta>0$, using
Proposition~\ref{prop:liouv-dynamics} with $\tfrac12-\alpha\in
B\left(\tfrac12+\vartheta,\zeta\right)$, and the fact that
$U(\theta)\Omega=\Omega$ for all $\theta\in\CC$, we can write for all
$0<|\lambda|<\Lambda$ and $\alpha\in\,]-\vartheta,1+\vartheta[$,
$$
\fF_{\omega,t}^\ttm(\alpha)
=\e^{\i t\cE(\lambda,\tfrac12-\alpha)} \left(\langle\Omega,\cQ_{\lambda,\frac12-\alpha}(\i r)\Omega\rangle
+O(\e^{-\lambda^2\epsilon t})\right)
$$
as $t\uparrow\infty$. If moreover
\beq\label{eq:nonzero-residue}
\langle\Omega,\cQ_{\lambda,\frac12-\alpha}(\i r)\Omega\rangle\neq0,
\eeq
then
\beq\label{eq:pref-ttm}
\bF_\omega^\ttm(\alpha)=\lim_{t\to\infty}\frac1t\log\fF_{\omega,t}^\ttm(\alpha)
=\i\cE\left(\lambda,\tfrac12-\alpha\right),
\eeq
and the map $\alpha\mapsto \bF_{\omega}^\ttm (\alpha)$ is analytic by
Proposition~\ref{prop:quasienergyspectrum}. We proceed to
prove \eqref{eq:nonzero-residue}.

Denote by $P_{\lambda,\alpha}$ the spectral projection of the quasi-energy
operator $\Sigma(\lambda,\alpha)$ associated to its eigenvalue
$\cE(\lambda,\alpha)$. The map $(\lambda,\alpha)\mapsto P_{\lambda,\alpha}$ is
analytic in each variable separately for $0<|\lambda|<\Lambda$ and $\alpha\in
B(\vartheta,\zeta)$. Since $\cE(\lambda,\alpha)$ is actually an eigenvalue of
$\Sigma_0(\lambda,\alpha)$, see Proposition~\ref{prop:quasienergyspectrum},
$P_{\lambda,\alpha}$ is also the spectral projection of
$\lambda^{-2}\Sigma_0(\lambda,\alpha)=\Sigma_0^{(2)}(\alpha)+O(\lambda)$ for the
eigenvalue $\lambda^{-2}\cE(\lambda,\alpha)=\cE^{(2)}(\alpha)+O(\lambda)$. It
follows that $\lambda\mapsto P_{\lambda,\alpha}$ extends analytically to
$\lambda=0$ where $P_{0,\alpha}=|X_\alpha\rangle\langle Y_\alpha|$ is the
spectral projection of $\Sigma_0^{(2)}(\alpha)$ for the eigenvalue
$\cE^{(2)}(\alpha)$, see Lemma~\ref{lem:lsospectrum}. Recall that
$X_\alpha,Y_\alpha$ are positive definite for $\alpha\in\RR$.

Using~\eqref{def:quasienergy} we then have
\[
\cQ_{\lambda,\alpha}(\i r)
=S_{\lambda,\alpha,0}(\i r)^{-1}P_{\lambda,\alpha}S_{\lambda,\alpha,0}(\i r),
\]
so the map $(\lambda,\alpha)\mapsto\cQ_{\lambda,\alpha}(\i r)$ is analytic
with $\cQ_{0,\alpha}(\i r)=P_{0,\alpha}$ (recall that $S_{0,\alpha,0}(\i r)$
is the identity). In particular, we have
\[
\langle\Omega, \cQ_{0,\alpha}(\i r)\Omega\rangle
=\langle\Omega_\sS,X_\alpha\rangle\langle Y_\alpha,\Omega_\sS\rangle
=\frac1N\tr(X_\alpha)\tr(Y_\alpha)>0
\]
for $\alpha$ real. By possibly making $\Lambda$ and $\zeta$ smaller
we derive that~\eqref{eq:nonzero-residue}, hence~\eqref{eq:pref-ttm},
holds for $|\lambda|<\Lambda$ and $\frac12-\alpha\in B\left(\frac12+\vartheta,\zeta\right)$.

Finally, the 2TMEP part of the PREF with respect to the NESS $\omega_+$, {\sl i.e.,}
$$
\bF_{\omega_+}^\ttm(\alpha)=\lim_{t\to\infty}\frac1t\log\fF_{\omega_+,t}^\ttm(\alpha)
=\i\cE\left(\lambda,\tfrac12-\alpha\right),
$$
follows from Theorem~\ref{prelim-thm}(3).

\ber Recall that $\fF_{\omega,t}^\ttm(\alpha)>0$ for $\alpha\in\RR$,
see e.g.~\eqref{def:ttm-measure}. This gives  that~\eqref{eq:pref-ttm}
implies that $\cE(\lambda,\alpha)$ is purely imaginary for
$\alpha\in\,]-\vartheta-1/2, \vartheta +1/2[$.
\eer

\ber Since $\fF_{\omega,t}^\ttm(0)=1$ for all $t$ we have
$\cE\left(\lambda,\frac12\right)=0$.
\eer

\subsection{QPSC part of the PREF}
\label{ssec:proof-qpsc}

Starting with the representation of Proposition~\ref{prop:liouv-functionals}(2)
we have, for all $\lambda,\alpha\in\CC$,
\[
\fF_{\omega_T,t}^\qpsc(\alpha)
=\langle\Omega,\e^{\i T\cL_{\lambda,1/2}}[D\omega_{-t}:D\omega]_\alpha\Omega\rangle.
\]
Corollary~\ref{coro:analytic-vectors} guarantees that
$[D\omega_{-t}:D\omega]_\alpha\Omega\in D$ so we can invoke
Proposition~\ref{prop:liouv-dynamics}. Since $\cE(\lambda,\tfrac12)=0$, we
obtain that for some $\Lambda>0$, all $0<|\lambda|<\Lambda$, and all $\alpha \in
\CC$,
\[
\fF_{\omega_+,t}^\qpsc(\alpha)
=\lim_{T\to\infty}\langle\Omega,\e^{\i T\cL_{\lambda,1/2}}[D\omega_{-t}:D\omega]_\alpha\Omega\rangle
=\langle\Omega,\cQ_{\lambda,\frac12}(\i r)U(\i r)[D\omega_{-t}:D\omega]_\alpha\Omega\rangle,
\]
where we again used the fact that $U(-\i r)\Omega=\Omega$.
Proposition~\ref{prop:liouv-functionals} leads to
\[
[D\omega_{-t}:D\omega]_\alpha\Omega=\e^{\i t\cL_{\lambda,1/2-\alpha}}\Omega,
\]
for all $\lambda\in\RR$ and $\alpha\in\CC$. Using again Corollary~\ref{coro:analytic-vectors},
the fact $U(\i r)\Omega=\Omega$ and invoking Lemma \ref{lem:semigroup-anlytic-continuation}
we further have
\[
\fF_{\omega_+,t}^\qpsc(\alpha)
=\langle\Omega, \cQ_{\lambda,\frac12}(\i r)\e^{\i t\cL_{\lambda,1/2-\alpha}(\i r)}\Omega\rangle.
\]
Using Proposition~\ref{prop:deformed-liouv-dynamics}, we hence get
\beq\label{eq:ness-qpsc}
\fF_{\omega_+,t}^\qpsc(\alpha)
=\e^{\i t\cE(\lambda,\frac12-\alpha)} \left(\langle\Omega,
\cQ_{\lambda,\frac12}(\i r)\cQ_{\lambda,\tfrac12-\alpha}(\i r)\Omega\rangle
+O(\e^{-\lambda^2\epsilon t})\right),
\eeq
so that
\[
\lim_{t\to\infty}\frac1t\log\fF_{\omega_+,t}^\qpsc(\alpha)
=\i\cE\left(\lambda,\tfrac12-\alpha\right)
\]
follows provided
\[
\langle\Omega,\cQ_{\lambda,\frac12}(\i r)\cQ_{\lambda,\frac12-\alpha}(\i r)\Omega\rangle\neq0.
\]
That for a given $\vartheta>0$ one can find $\Lambda>0$ and $\zeta>0$ such that
the latter identity holds for $|\lambda|<\Lambda$  and
$\tfrac12-\alpha\in B(\tfrac12+\vartheta,\zeta)$ is now deduced by following
the proof of the related relation~\eqref{eq:nonzero-residue} given in
Section~\ref{ssec:proof-ttm}.

\ber\label{rem:subseq} When $\lambda=0$, we actually have
\[
\langle\Omega,\cQ_{0,\frac12}(\i r)\cQ_{0,\frac12-\alpha}(\i r)\Omega\rangle
=\langle\Omega_\sS,X_{\frac12}\rangle\langle Y_{\frac12},X_{\frac12-\alpha}\rangle
\langle Y_{\frac12-\alpha},\Omega_\sS\rangle>0.
\]
This ensures that the logarithm of the complex valued quantity
$\fF_{\omega_+,t}^\qpsc(\alpha)$ is indeed well-defined for $\lambda$
small and $t$ large.
\eer

\subsection{EAST part of the PREF}
\label{ssec:proof-ancilla}

The proof is completely parallel to the one of the previous section. The same
reasoning starting from the representation given in
Proposition~\ref{prop:liouv-functionals}(3) gives that, for some $\Lambda >0$,
all $0<|\lambda|<\Lambda$ and all $\alpha\in\CC$,
\begin{align*}
\fF_{\omega_+,t}^\anc(\alpha)
&=\lim_{T\to\infty}\langle\Omega,\e^{\i T\cL_{\lambda,1/2}}
[D\omega_{-t}:D\omega]^\ast_{\frac{\bar\alpha}2}[D\omega_{-t}:D\omega]_{\frac{\alpha}2}\Omega\rangle\\[2mm]
&=\langle\Omega,\cQ_{\lambda,\frac12}(\i r)U(\i r)
[D\omega_{-t}:D\omega]^\ast_{\frac{\bar\alpha}2}[D\omega_{-t}:D\omega]_{\frac{\alpha}2}\Omega\rangle\\[2mm]
&=\langle\Omega,\cQ_{\lambda,\frac12}(\i r)U(\i r)\e^{it\widehat{\cL}_{\lambda,\alpha}}\Omega\rangle\\[2mm]
&=\langle\Omega,\cQ_{\lambda,\frac12}(\i r)\e^{it\widehat\cL_{\lambda,\alpha}(i r)}\Omega\rangle\\[2mm]
&=\e^{\i t\widehat{\cE}(\lambda,\alpha)}\left(
\langle\Omega,\cQ_{\lambda,\frac12}(\i r)\widehat{\cQ}_{\lambda,\alpha}(\i r)\Omega\rangle
+O(\e^{-\lambda^2\epsilon t})\right),
\end{align*}
where we have also used~\eqref{eq:semigroup-deformation}
and~\eqref{eq:deformed-liouv-dynamics2}. By exactly the same argument as in
Section~\ref{ssec:proof-ttm} one can find $\Lambda >0$ such that for
$|\lambda|<\Lambda$ and $\alpha\in B(\vartheta,\zeta)$ one has
\[
\langle\Omega,\cQ_{\lambda,\frac12}(\i r)\widehat{\cQ}_{\lambda,\alpha}(\i r)\Omega\rangle\neq0,
\]
so that
\[
\bF_{\omega_+}^\anc(\alpha)=\lim_{t\to\infty}\frac1t\log\fF_{\omega_+,t}^\anc(\alpha) =\i\widehat{\cE}\left(\lambda,\alpha\right).
\]

It remains to prove that this limit coincides with the one of the 2TMEP and QPSC functionals, {\sl i.e.,} that
\beq\label{eq:E-hatE}
\widehat{\cE}(\lambda,\alpha)=\cE\left(\lambda,\tfrac12-\alpha\right).
\eeq
Recall that $\fF_{\omega,t}^\anc(\alpha)=\fF_{\omega,t}^\ttm(\alpha)$, see~\eqref{eq:functionals-equality}. By Proposition~\ref{prop:liouv-functionals}(3), we have
\[
\fF_{\omega,t}^\anc(\alpha) =\langle\Omega,\e^{it\widehat{\cL}_{\lambda,\alpha}}\Omega\rangle,
\]
so, using~\eqref{eq:liouv-dynamics2} we further get
\[
\lim_{t\to\infty}\frac1t\log\fF_{\omega,t}^\anc(\alpha) =\i \widehat\cE\left(\lambda, \alpha\right).
\]
Combined with~\eqref{eq:pref-ttm} this proves~\eqref{eq:E-hatE}.

\subsection{Nonvanishing of $\bF$}
\label{ssec:nonvanishing}

It remains to establish the assertion dealing with the non-vanishing of $\bF$.
As mentioned in Remark~\ref{rem:follow}, the convexity and analyticity of $\bF$,
combined with the symmetry $\bF(0)=\bF(1)=0$, ensure that either $\bF$ is
identically vanishing on $]-\vartheta,1+\vartheta[$ or is strictly convex. If it
is strictly convex, the symmetry also guarantees that $0$ is not a minimum so
that $\bF$ is identically vanishing if and only if
$$
\partial_{\alpha}\bF(\alpha)|_{\alpha=0}=0.
$$
Now, using~\eqref{ep-eq} and~\eqref{eq:functional-derivative} we have
\[
\partial_{\alpha}\fF_{\omega,t}^\ttm(\alpha)\bigg|_{\alpha=0}=-\int_0^t\omega_s(\sigma)\d s.
\]
This identity and convexity (see e.g.~\eqref{def:ttm-measure}) give
\[
\partial_{\alpha}\bF(\alpha)|_{\alpha=0}=-\omega_+(\sigma).
\]
It thus remains to prove that $\omega_+(\sigma)=0$ if and only if $\beta_1=\cdots=\beta_M=\beta$.

It is  proven in~\cite[Theorems~1.3--1.4]{Jaksic2002a}
and~\cite[Theorem~1.15]{Jaksic2006a} that if $\beta_i\not=\beta_j$ for some $i,
j$, then under the assumptions of Theorem~\ref{main-thm-sfm} there exists
$\Lambda >0$ such that for $0<|\lambda|<\Lambda$, $\omega_+(\sigma)>0$. On the
other hand, if $\beta_1=\cdots=\beta_M=\beta$ then $\omega_+$ is a
$(\tau_\lambda,\beta)$-KMS state and in particular $\omega_+\in\cN$.
By~\cite[Theorem~1.3]{Jaksic2003}, $\omega_+(\sigma)=0$.

\subsection{The simplest spin--fermion model}
\label{ssec:simple}

The spectrum of $\cL_\sS$ is $\{-2,0,2\}$, the eigenvalue $0$ having
multiplicity $2$. Using~\eqref{eq:lso} one can compute explicitly
$\Sigma_e^{(2)}(\alpha)$. For $\alpha=0$ and $\alpha=\frac12$ this was done
in~\cite{Jaksic2002a}, and one can then use~\eqref{eq:lso-similarity} to obtain
$$
\Sigma_0^{(2)}(\alpha) =\i\pi\sum_{j=1}^M\frac{\|\tilde{f}_j(2)\|^2_\fH}{2\cosh\beta_j}
\begin{bmatrix}
\e^{\beta_j}& -\e^{2\alpha\beta_j}\\
-\e^{-2\alpha\beta_j}&\e^{-\beta_j}
\end{bmatrix},
$$
while $\Sigma_{\pm2}^{(2)}(\alpha)$ are scalars, which turn out to be
independent of $\alpha$, and are given by
\[
\Sigma_{\pm2}^{(2)}(\alpha)
=\frac12\sum_{j=1}^M\left( \mp\mathrm{PV}\!\!\int_\RR
\frac{\|\tilde f_j(r)\|^2_\fH}{r-2}\d r+\i\pi\|\tilde f_j(2)\|^2_{\fH} \right),
\]
where $\mathrm{PV}$ stands for Cauchy's Principal Value.

The eigenvalues of $\Sigma_0^{(2)}(\alpha)$ are
\[
\begin{split}
E_{0\pm}^{(2)}(\alpha)=\i\frac{\pi}2\left( \sum_{j=1}^M\right.&\|\tilde f_j(2)\|^2_{\fH_j}\\
&\left.\pm\sqrt{\sum_{j,k=1}^M \left(\tanh(\beta_j)\tanh (\beta_k)
+\frac{\cosh(2(\beta_j-\beta_k)\alpha)}{\cosh(\beta_j)\cosh(\beta_k)}\right)
\|\tilde f_j(2)\|^2_{\fH}\|\tilde f_k(2)\|^2_{\fH}}\,\right).
\end{split}
\]
Obviously, $E_{0-}^{(2)}(\alpha)$ has the smallest imaginary part, so that
$$
\cE^{(2)}(\alpha)=E_{0-}^{(2)}(\alpha),
$$
and~\eqref{eq:simplesf-functional} follows from~\eqref{eq:pref-ttm}.

\subsection{Comparison with the general scheme of~\cite{Benoist2024}}
\label{ssec:remarks}

Although our analysis of the $\alpha$-Liouvilleans mostly follows the abstract
scheme given in~\cite{Benoist2024}, the structural properties of the
spin-fermion model allow to simplify certain steps. We have for example used
Propostion~\ref{prop:liouv-functionals}(3) to analyze the ancilla part of PREF,
therefore relying on the variant {\bf~(Deform2A)} of the general scheme. Also,
the regularity of the map $\alpha\mapsto\cE(\lambda,\alpha)$, hence of
$\alpha\mapsto\bF(\alpha)$, is here a consequence of regular perturbation theory
that allows us to bypass Assumption~{\bf (Deform3)}, see also Remark 2 after
Theorem 4.5 in \cite{Benoist2024}.

\bibliographystyle{capalpha}
\bibliography{SF}
\end{document}